\documentclass[12pt, epsfig]{article}
\usepackage{amsmath,amsfonts,amssymb,amsthm,amstext,amscd,eucal,graphicx, xcolor}
\usepackage[all]{xy}
\usepackage{dsfont}
\usepackage{amsmath}
\usepackage{slashed,ccaption}

\usepackage[unicode=true,
 bookmarks=false,
 breaklinks=false,pdfborder={0 0 1},backref=false,colorlinks=true]
 {hyperref}
\hypersetup{
 pdfauthor={Hamid Afshar, Daniel Grumiller, Shahin Sheikh-Jabbari and Hossein Yavartanoo},
 linkcolor=blue,urlcolor=magenta,filecolor=green,citecolor=red,pdfstartview=FitV,bookmarksopen=true}

\makeatletter \@addtoreset{equation}{section}

\makeatletter\renewcommand\section{\@startsection {section}{1}{\z@}%
                                   {-3.5ex \@plus -1ex \@minus -.2ex}
                                   {2.3ex \@plus.2ex}%
                                   {\normalfont\large\bfseries}}
\renewcommand\subsection{\@startsection{subsection}{2}{\z@}%
                                     {-3.25ex\@plus -1ex \@minus -.2ex}%
                                     {1.5ex \@plus .2ex}%
                                     {\normalfont\bfseries}}

\makeatletter
\DeclareFontFamily{OMX}{MnSymbolE}{}
\DeclareSymbolFont{MnLargeSymbols}{OMX}{MnSymbolE}{m}{n}
\SetSymbolFont{MnLargeSymbols}{bold}{OMX}{MnSymbolE}{b}{n}
\DeclareFontShape{OMX}{MnSymbolE}{m}{n}{
    <-6>  MnSymbolE5
   <6-7>  MnSymbolE6
   <7-8>  MnSymbolE7
   <8-9>  MnSymbolE8
   <9-10> MnSymbolE9
  <10-12> MnSymbolE10
  <12->   MnSymbolE12
}{}
\DeclareFontShape{OMX}{MnSymbolE}{b}{n}{
    <-6>  MnSymbolE-Bold5
   <6-7>  MnSymbolE-Bold6
   <7-8>  MnSymbolE-Bold7
   <8-9>  MnSymbolE-Bold8
   <9-10> MnSymbolE-Bold9
  <10-12> MnSymbolE-Bold10
  <12->   MnSymbolE-Bold12
}{}

\let\llangle\@undefined
\let\rrangle\@undefined
\DeclareMathDelimiter{\llangle}{\mathopen}%
                     {MnLargeSymbols}{'164}{MnLargeSymbols}{'164}
\DeclareMathDelimiter{\rrangle}{\mathclose}%
                     {MnLargeSymbols}{'171}{MnLargeSymbols}{'171}
\makeatletter
\newcommand{\be}{\begin{equation}}
\newcommand{\ee}{\end{equation}}
\newcommand{\bea}{\begin{eqnarray}}
\newcommand{\eea}{\end{eqnarray}}
\newcommand{\bse}{\begin{subequations}}
\newcommand{\ese}{\end{subequations}}
\newcommand{\beqa}{\begin{eqnarray}}
\newcommand{\eeqa}{\end{eqnarray}}
\newcommand{\beqar}{\begin{eqnarray*}}
\newcommand{\eeqar}{\end{eqnarray*}}
\newcommand{\bi}{\begin{itemize}}
\newcommand{\ei}{\end{itemize}}
\newcommand{\bn}{\begin{enumerate}}
\newcommand{\en}{\end{enumerate}}

\newcommand{\ba}{\begin{array}}
\newcommand{\ea}{\end{array}}
\newcommand{\bc}{\begin{center}}
\newcommand{\ec}{\end{center}}

\newcommand{\nn}{\nonumber}

\newcommand{\cJ}{\mathcal J}
\newcommand{\cW}{\mathcal W}
\newcommand{\bcJ}{\boldsymbol{\mathcal J}}
\newcommand{\bcW}{\boldsymbol{\mathcal W}}
\newcommand{\bcL}{\boldsymbol{\mathcal L}}
\newcommand{\bJ}{\boldsymbol{J}}
\newcommand{\bh}{\boldsymbol{h}}

\newcommand{\lbL}{\mbox{\large{$\boldsymbol{L}$}}}
\newcommand{\bL}{\boldsymbol{L}}
\newcommand{\bN}{\boldsymbol{N}}

\newcommand{\cB}{\mathcal B}

\newcommand{\cH}{\mathcal H}

\newcommand{\HBH}{\mathcal H_{\textrm{\tiny BTZ}}}

\newcommand{\HCG}{\mathcal H_{\textrm{\tiny CG}}}
\newcommand{\HConic}{\mathcal H_{\textrm{\tiny Conic}}}
\newcommand{\Hglobal}{\mathcal H_{\textrm{\tiny gAdS}}}

\def\nn{\nonumber}

\DeclareMathOperator{\extdm}{d}
\newcommand{\extd}{\extdm \!}

\newcommand{\vp}{\varphi}

\definecolor{darkgreen}{rgb}{0,0.3,0}
\definecolor{darkblue}{rgb}{0,0,0.3}
\definecolor{darkred}{rgb}{0.7,0,0}

\newcommand{\old}[1]{}

\newcommand{\eq}[2]{\begin{equation} #1 \label{#2} \end{equation}}

\begin{document}
\renewcommand{\baselinestretch}{1.2}  

\begin{titlepage}

\begin{flushright}\vspace{-3cm}
{
IPM/P-2017/017  \\
\today }\end{flushright}
\vspace{-1cm}

\newcommand{\mytitle}{Horizon fluff, semi-classical black hole microstates}
\newcommand{\mysubtitle}{\centerline{Log-corrections to BTZ entropy and  black
hole/particle correspondence}}

\bigskip

\begin{center}
\Large{\bf{\hspace*{-.5cm} \mytitle}}\\
\medskip
\large{\bf{\mysubtitle}}

\bigskip\bigskip

\large{\bf{H.~Afshar\footnote{e-mail:~\href{mailto:afshar@ipm.ir}{afshar@ipm.ir}}$^{; ab}$, D.~Grumiller\footnote{e-mail:~\href{grumil@hep.itp.tuwien.ac.at}{grumil@hep.itp.tuwien.ac.at}}$^{; c}$, M.M.~Sheikh-Jabbari\footnote{e-mail:~\href{jabbari@theory.ipm.ac.ir}{jabbari@theory.ipm.ac.ir}}$^{; a}$, H.~Yavartanoo\footnote{e-mail:~\href{yavar@itp.ac.cn}{yavar@itp.ac.cn}}$^{; d}$
}}
\\

\vspace{5mm}
\normalsize
\bigskip

{$^a$ \it School of Physics, Institute for Research in Fundamental
Sciences (IPM),\\ P.O.Box 19395-5531, Tehran, Iran}\\

\smallskip

{$^b$ \it Theory Group, Physics Department, CERN, CH-1211 Geneva 23,
Switzerland}

\smallskip

{$^c$ \it Institute for Theoretical Physics, TU Wien, Wiedner Hauptstr.~8, A-1040 Vienna, Austria
 }
 
\smallskip

{$^d$ \it  State Key Laboratory of Theoretical Physics, Institute of Theoretical Physics,
Chinese Academy of Sciences, Beijing 100190, China.
 }\\

\date{\today}

\end{center}
\setcounter{footnote}{0}

\bigskip

\begin{abstract}

\noindent
According to the \emph{horizon fluff} proposal microstates of a generic black hole belong to a certain subset of near horizon soft hairs that cannot be extended beyond the near horizon region. In \cite{Afshar:2016uax,Sheikh-Jabbari:2016npa} it was shown how the horizon fluff proposal works for AdS$_3$ black holes. In this work we clarify further this picture by showing that BTZ black hole microstates are in general among the {coherent states in the Hilbert space associated with  conic spaces or their Virasoro descendants}, provided we impose a (Bohr-type) quantization condition on the angular deficit. Thus BTZ black holes may be viewed as condensates (or solitonic states) of AdS$_3$ particles. We provide canonical and microcanonical descriptions of the statistical mechanical system associated with BTZ black holes and their microstates, and relate them. As a further non-trivial check we show the horizon fluff proposal correctly reproduces the expected logarithmic corrections to the BTZ entropy. 
\end{abstract}



\end{titlepage}
\tableofcontents

\section{Motivations and introduction}

\subsection{Introductory remarks}

Black hole entropy has some surprisingly universal properties that are accessible through semi-classical considerations. The most prominent one is the thermodynamic description, and among the thermodynamic quantities is the black hole entropy, which unlike other black hole thermodynamic quantities is observer independent. The black hole entropy in Einstein gravity theories is given by the Bekenstein--Hawking entropy law  \cite{Bekenstein:1973ur, Hawking:1975sw},  
\be
S_{\textrm{\tiny BH}} =\frac{A}{4G}
\label{eq:BH}
\ee
according to which the black hole entropy $S_{\textrm{\tiny BH}}$ is a quarter of the area $A$ of the black hole horizon in units of Newton's constant $G$. The Bekenstein--Hawking  law can be derived e.g.~in the saddle point approximation to the Euclidean path integral of Einstein gravity \cite{Gibbons:1976ue} or as the conserved charge associated with the Killing horizon generating vector field in the Einstein--Hilbert theory \cite{Wald:1993nt}. 

The area law generically receives  subleading logarithmic corrections
\be
S = S_{\textrm{\tiny BH}} - N_{\textrm{\tiny log}}\,\ln S_{\textrm{\tiny BH}} + {\cal O}(1)
\label{eq:logS}
\ee
which are specified by a numerical coefficient $N_{\textrm{\tiny log}}$ that is accessible by semi-classical means and does not depend on the details of the UV-completion of general relativity (GR), see \cite{Sen:2011ba,Sen:2012dw,Carlip:2000nv, Loran:2010bd, Keeler:2014bra,Charles:2015eha} and references therein. Since the calculations involved in deriving \eqref{eq:BH} and \eqref{eq:logS} require only theories and methods that have been tested to high accuracy experimentally, namely GR and perturbative quantum field theory, these formulas are the closest template to a positive experimental result we have for gravity beyond classical GR.\footnote{We inserted the attribute `positive', since there are negative experimental results that rule out already some speculations about quantum gravity, for instance the absence of certain modified dispersion relations or Lorentz violations at the Planck scale \cite{Ackermann:2009aa}.}

During the past year an alternative representation of black hole entropy has emerged that requires also only the validity of the semi-classical approximation, see \cite{Afshar:2016wfy} and its followups. The Bekenstein--Hawking formula \eqref{eq:BH} is replaced by
\eq{
S = 2\pi\,\big({\tt J}_0^+ + {\tt J}_0^-\big)
}{eq:intro1}
where ${\tt J}_0^\pm$ are zero-mode charges in some `near horizon symmetry algebra', which we shall review in later sections. Originally, the result \eqref{eq:intro1} was derived for Einstein gravity in three dimensional anti-de~Sitter space (AdS$_3$) \cite{Afshar:2016wfy}. During the past year it was shown that the same result applies to gravitational theories in AdS$_3$ with higher derivative interactions \cite{Setare:2016vhy} or massless higher spin interactions \cite{Grumiller:2016kcp}, and to three-dmensional flat space Einstein gravity \cite{Afshar:2016kjj} and flat space higher spin gravity \cite{Ammon:2017vwt}. 

At least in three dimensions, the main focus of the present work, the result \eqref{eq:intro1} is even more universal than the Bekenstein--Hawking law, since it applies also in cases where Bekenstein--Hawking fails (such as higher derivative or higher spin theories). Moreover, while \eqref{eq:intro1} was derived through analysis at the classical level, its Cardy-like form suggests a simple dual field theoretical interpretation. This provides a first hint that (semi-)classical considerations phrased in a `near horizon picture' (we shall be more precise below what we mean by this notion) may provide insights into black hole microstates. Before continuing this line of reasoning we need another semi-classical ingredient, asymptotic symmetries and their near horizon counterpart, that we review in the next couple of paragraphs.

The flip side of the success of semi-classical GR in explaining black hole entropy is the information paradox, which seems impossible to resolve semi-classically: Since black holes have an entropy and a temperature, they generically Hawking-radiate, which can lead to their evaporation \cite{Hawking:1974rv, Iyer:1994ys, Wald:1993nt}. This implies that physical information could permanently disappear in the black hole evaporation, allowing many physical states to devolve into the same thermal state, which violates unitarity \cite{Hawking:1976ra}.

To avoid loss of unitarity, standard arguments of statistical mechanics suggest the existence of black hole microstates with a large number of degeneracy. However, black hole uniqueness and no hair theorems \cite{Ruffini:1975ne, Chrusciel:1994sn, Bekenstein:1996pn} rule out simple realization of these microstates as locally distinct solutions to Einstein gravity. This is why many attempts to construct black hole microstates rely on some UV completion of GR, such as string theory \cite{Strominger:1996sh}. However, in a UV completed theory the universality of the semi-classical results \eqref{eq:BH}-\eqref{eq:intro1} is not obvious, which may suggest that there could be a loophole in these considerations and it is not implausible  to construct microstates  semi-classically in a suitable formulation of the theory. Indeed, it has been stressed in various recent publications \cite{Hawking:2016msc, Sheikh-Jabbari:2016lzm} that there is a loophole to the uniqueness results, and one should refine the strict statement of the equivalence principle allowing a very particular set of diffeomorphic geometries to be physically distinct.

Developments in GR which started in mid-1960's led to the conclusion that, like in any field theory, to fully define dynamics besides the equations of motion we need to specify boundary behavior of the fields. In contrast to many field theory setups ``natural'' boundary conditions, where all fields vanish sufficiently fast near the (asymptotic) boundary, are not available in gravity since the field in question is the metric, which cannot vanish in a region where classical gravity is a good approximation.

In particular, for four-dimensional asymptotically flat solutions to Einstein gravity, it was noted that there are diffeomorphic geometries which differ by their boundary behavior and that the conserved charges associated with the diffeomorphisms relating these geometries form an infinite dimensional algebra --- the asymptotic symmetry algebra --- now known as BMS$_4$ algebra \cite{Bondi:1962, Sachs:1962}. This analysis was examined for many other geometries in various dimensions, in particular for solutions to AdS$_3$ Einstein gravity. In their seminal paper \cite{Brown:1986nw},  Brown and Henneaux showed that by imposing appropriate asymptotic behavior of the metric near the AdS$_3$ boundary one finds the two-dimensional conformal algebra as asymptotic symmetry algebra, which provided a precursor for AdS$_3$/CFT$_2$. Further analysis  has then revealed that there is the possibility of imposing different boundary conditions than the Brown--Henneaux ones, with different symmetry algebras. These boundary conditions could be more restrictive than the Brown--Henneaux ones, e.g.~\cite{Balasubramanian:2009bg}, or less restrictive, e.g.~\cite{Troessaert:2013fma, Grumiller:2016pqb}, or simply different, e.g.~\cite{Afshar:2016wfy, Compere:2013bya, Avery:2013dja, Donnay:2015abr, Afshar:2015wjm,  Donnay:2016ejv, Majhi:2017fua}. 

\subsection{Black hole microstates (horizon fluff) as near horizon soft hairs}

In our work we will particularly exploit the boundary conditions of \cite{Afshar:2016wfy}. While they may be formulated as asymptotic conditions on the metric, they take their simplest and most natural form in a near horizon expansion of the metric, which is one of the reasons we refer to them as `near horizon boundary conditions' and  call the associated  symmetry algebra `near horizon symmetry algebra', in order to distinguish it from other asymptotic symmetry algebras (e.g. that of Brown and Henneaux). We note in passing that the usual analysis of gauge systems with suitable boundary conditions in the presence of asymptotic boundaries eventually leads to an expression for the boundary charges and the ensuing asymptotic or symplectic symmetry 
algebra \cite{Compere:2015knw}.

One of the main ingredients that we exploit is the generality of the near horizon algebra and the simplicity of the semi-classical entropy \eqref{eq:intro1} when expressed in terms of zero mode charges of the near horizon symmetry algebra. 
The existence of these asymptotic symmetry algebras calls for a revision of strict general covariance and the equivalence principle \cite{Sheikh-Jabbari:2016lzm}.\footnote{%
By ``strict general covariance'' we mean the statement that any two geometries locally diffeomorphic to each other are physically equivalent. The BMS and Brown--Henneaux analyses provide counterexamples to strict general covariance, since locally diffeomorphic geometries like AdS$_3$ and BTZ black holes \cite{Banados:1992wn, Banados:1992gq} are physically different.
} It may also open the venue for identifying  black hole microstates \cite{Afshar:2016uax, Sheikh-Jabbari:2016npa} and possibly resolving other issues associated with black holes, like the information paradox \cite{Hawking:2016msc, Hawking:2016sgy, Compere:2016jwb, Compere:2016hzt, Compere:2016gwf} (see however, \cite{Shahin:2016, Mirbabayi:2016axw, Gabai:2016kuf}). A prominent entity in recent discussions is ``soft hair'' \cite{Hawking:2016msc}, referring to zero energy excitations with non-trivial canonical charges. One should also note that in a curved background due to the existence of various possible observers with different notions of time, ``softness'' is an observer dependent notion. In particular, here we are dealing with ``near horizon soft hairs'' which differ from the ``asymptotic soft hairs'' discussed in \cite{Hawking:2016msc}. 

One of the immediate problems with the idea \cite{Hawking:2015qqa} that soft hair could account for the Bekenstein--Hawking entropy is that there seems to be no upper bound on the amount of soft hair that a black hole can carry, precisely because of its softness. The ``horizon fluff'' proposal improves on the soft hair idea by providing a natural cut-off on the near horizon soft hair spectrum \cite{Afshar:2016uax, Sheikh-Jabbari:2016npa}. Explaining how this works in detail is one of the main goals of this paper, where we intend to study further and scrutinize the horizon fluff proposal. 

The first step towards horizon fluff was performed in \cite{Afshar:2016uax}, where a simple model for identifying black hole microstates was proposed. A certain class of near horizon soft hairs are associated with a non-extremal BTZ black hole solution \cite{Banados:1992wn, Banados:1992gq} or its conformal descendants \cite{Banados:1998gg,Sheikh-Jabbari:2014nya,Sheikh-Jabbari:2016unm}. These states are identified as the black hole microstates and were called horizon fluff. One particular outcome was that the soft hair spectrum was cut off, and the degeneracy of the horizon fluff reproduced in the classical limit the Bekenstein--Hawking entropy \eqref{eq:BH} in its near horizon formulation \eqref{eq:intro1}. The analysis was extended for all non-extremal black holes in the class of Ba\~nados geometries \cite{Banados:1998gg} in \cite{Sheikh-Jabbari:2016npa}, where it was shown that the resulting entropy, as expected, is an invariant of Virasoro coajoint orbits.

\subsection{Main results and organization of the paper}

In the present work we employ the same general setup, but study black hole microstates from a slightly different angle and refine the horizon fluff proposal. We summarize now the main results that we shall derive in later sections. 

We start in section \ref{constant-rep} by providing `canonical' and `microcanoncial' descriptions for locally AdS$_3$ geometries with certain boundary conditions and an explicit map between the two. The former is characterized by a function $\Phi$ with periodicity property $\Phi(\phi+2\pi)=\Phi(\phi) \pm 2\pi J_0$, while the latter is characterized by a function $h$ with $h(\phi+2\pi)=h(\phi)+2\pi$ and $h'>0$. Their relation is given by
\eq{
\Phi = \int^\phi J(\phi)= \pm J_0 h - \frac12\,\ln h^\prime\,.
}{eq:sum1}
The two signs in the equation above show that there are two physically equivalent solutions for $\Phi$ for each $h$, related by $J_0\to -J_0$. The quantity $J_0$ turns out to be real for BTZ solutions and imaginary for conical defects and global AdS (for notational simplicity we mostly suppress the labels referring to left- and right-moving sectors, but we note that we have two essentially identical copies of all quantities, e.g.~$J_0^\pm$).

We then investigate the Hilbert space associated with the canonical (symplectic) symmetry algebra, which consists of two $\hat u(1)_k$ current algebras $\boldsymbol{J}_n$, 
\be\label{J-c-intro}
{[\bJ_n, \bJ_m]=\frac{c}{12} n\delta_{n+m,0},\qquad c=6k=\frac{3\ell}{2G},}
\ee
with $\ell$ being the AdS$_3$ radius and $G$ 3d Newton constant, and divide it into the black hole subspace ${\cal H}_{\textrm{\tiny BTZ}}$ and the remainder ${\cal H}_{\textrm{\tiny CG}}$. The former contains BTZ black holes corresponding to real $J_0$, while the latter is characterized by imaginary $J_0=i\nu/2$, $\nu\in(0,1]$ and thus describes conical defects and global AdS. To describe the subspace ${\cal H}_{\textrm{\tiny CG}}$ we introduce a primary field of weight one defined by
\eq{
{\cal W} = e^{-2\Phi}
}{eq:sum2}
for each value of $\nu$. An interesting consequence of our definition is that the $\cal W$-fields obey twisted periodicity conditions,
\eq{
{\cal W}^{\pm \nu}(\phi+2\pi) = e^{{\mp} 2\pi\nu i}\,{\cal W}^{\pm \nu}(\phi)
}{eq:sum3}
where $\nu$ (introduced above) is a state-independent quantity parametrizing the conical deficit. Using  consistency of the algebra of operators, we argue that the $\cal W$-fields should satisfy canonical quantization conditions to leading order in large $c$.

We then assume (and give arguments in favor of) two semi-classical quantization conditions: \begin{enumerate}
\item \vskip -3mm the central charge $c$ \eqref{J-c-intro} is quantized in positive integers; 
\item  \vskip -3mm the angular deficit is quantized in integers over $c$, \eq{
\nu=r/c \qquad\textrm{with}\; r=1,2,\dots,c\,.
}{eq:sum4}
\end{enumerate}
The second assumption, our ``Bohr-quantization'' of conical deficit angles, is perhaps the less trivial one, but appears to be justified by explicit string theory realizations. 

A nice technical aspect of our construction is that all Fourier modes of all ${\cal W}^\nu$-fields together can be reassembled into two $\hat u(1)$ current algebras $\boldsymbol{{\cal J}}_n$,
\be{
[\bcJ_n, \bcJ_m]=\frac{n}{2}\delta_{n+m,0},}
\ee
which is then used to reconstruct the ${\cal H}_{\textrm{\tiny CG}}$ in a different basis. This Hilbert space is spanned by the set of all states created by the action of $\bcJ_n$ current generators on the vacuum $|0\rangle$, defined through $\bcJ_n|0\rangle=0,\ n\geq 0$. States in this Hilbert space provide the basis of our black hole microstates, as described below. 

The main new conceptual ingredient as compared to our earlier works is that we propose a correspondence in the Hilbert space of the symmetry algebra of conserved charges, the Virasoro Hilbert space $\mathcal H_{\textrm{\tiny Vir}}$. This correspondence is a map between states in $\mathcal H_{\textrm{\tiny Vir}}$ in the hermitian and anti-hermitian representation of the symmetry algebra. 
The former includes BTZ black holes and their descendants and the latter has conic spaces, global AdS$_3$ and their descendants. 
In particular, we argue that the conic and global AdS$_3$ parts of the symmetry algebra  admit an alternative two-dimensional conformal field theory (CFT$_2$) free field description (described by $\bcJ_n$) and the correspondence is simply requiring these two descriptions to be equivalent. The key relation of our proposed black holes/particle correspondence to leading order in $c$ is given by
\eq{
\frac{1}{c}\sum_{p\in\mathbb{Z}}  \colon\!\bcJ_{nc-p}\bcJ_p\!\colon=in \bJ_n+ \frac{6}{c}\sum_{p\in\mathbb{Z}} \colon\!\bJ_{n-p}\bJ_p\!\colon
}{eq:sum5}
which relates the current algebra generators $\boldsymbol{{\cal J}}_n$ and  $\boldsymbol{J}_n$ and hence the corresponding Hilbert spaces.

Using this correspondence we map black hole states to a collection of conic spaces described through $\cJ_n$. This allows us to identify black hole microstates (in a weakly coupled dual description) as a gas of coherent states of particles on AdS$_3$. Alternatively, in our description a BTZ black hole is viewed as a condensate of these coherent states. Explicitly, the set of all BTZ microstates $|\cB(\{n_i^\pm\}); J^\pm_0\rangle$ with mass and angular momentum given by the parameters $\Delta_\pm = \tfrac12\,(\ell M\pm J) = \tfrac c6\,(J_0^\pm)^2$ is labelled by a set of integers $\{n_i^\pm\}$ and reads
\eq{
|\cB(\{n_i^\pm\}); J^\pm_0\rangle = \prod_{\{n_i^\pm>0\}} \!\!\!\!\big(\bcJ_{-n_i^+}^+ \cdot \bcJ_{-n_i^-}^-\big) |0\rangle \,,\quad \text{such that}\quad \sum n_i^\pm= c\Delta^\pm\,.
}{eq:sum6}
Counting the degeneracy of these states reproduces the Bekenstein--Hawking entropy \eqref{eq:BH}, \eqref{eq:intro1} and also gives the correct numerical coefficient $N_{\textrm{\tiny log}}=-3/2$ in the subleading logarithmic corrections \eqref{eq:logS}, after taking into account an additional log-term that arises when switching between canonical and microcanonical descriptions.

This  paper is organised as follows. In section \ref{constant-rep}, we review and construct the phase space of locally AdS$_3$ black holes and provide two descriptions for them, the canonical and the microcanonical one. These two are distinguished by which physical quantity is held fixed by the choice of boundary conditions on the allowed diffeomorphisms. In section \ref{sec-canonical-Hilbert}, we discuss symplectic symmetries and phase space for the canonical description and discuss the corresponding Hilbert space. In section \ref{sec:more-on-HCG} we present another free field CFT$_2$ description for the class of conic spaces and their Virasoro descendants. In section \ref{sec:two-desc}, we introduce the black hole/particle correspondence and present the horizon fluff proposal in light of this duality. We explicitly construct all the microstates and count them employing the Hardy--Ramanujan formula. Section \ref{sec:log-correction} contains our main new result for black hole entropy, where we show that the horizon fluff proposal correctly reproduces also the logarithmic corrections to the BTZ black hole entropy. For the latter we need to carefully analyze the canonical and microcanonical descriptions and the interplay between them. Section \ref{sec:discussion} is devoted to concluding remarks and insights, comparison with other approaches and possible future directions. The appendices contain supplementary material on the Ba\~nados geometries and the corresponding Virasoro Hilbert space (appendix \ref{app:A}), conic spaces (appendix \ref{app:Conic}), a discussion of hermitian conjugation and zero point energies (appendix \ref{appendix-D}) and a derivation of the Ba\~nados map \cite{Banados:1998wy} (appendix \ref{app:Banados-map}).

\subsection{Conventions}\label{convention}

Before starting let us mention some conventions. The theory we choose to study is Einstein gravity in three dimensions with negative cosmological constant, $\Lambda=-1/\ell^2$, parametrized by the AdS radius $\ell>0$. 

In Einstein--Hilbert--Palatini or, equivalently, Chern--Simons formulation the bulk action $I_{\textrm{\tiny EHP}}$ is given by \cite{Achucarro:1986vz, Witten:1988hc},
\eq{
I_{\textrm{\tiny EHP}} = \frac{k}{4\pi}\,\big(I_{\textrm{\tiny CS}}[A^+] - I_{\textrm{\tiny CS}}[A^-]\big),
}{eq:ICS}
where $A^\pm$ are $sl(2,\mathbb{R})$ connections, and $I_{\textrm{\tiny CS}}[A]=\int\llangle A\wedge \extd A + \tfrac23\,A\wedge A\wedge A\rrangle$ is the Chern--Simons 3-form integrated over some manifold that we assume to be topologically a (filled) cylinder or torus. $\llangle\rrangle$ denotes the associated bilinear form whose value here for $sl(2,\mathbb{R})$ is $\llangle L_1,L_{-1}\rrangle=-2\llangle L_0,L_0\rrangle=-1$.  The only coupling constant is the Chern--Simons level $k$ given in \eqref{J-c-intro}.
The metric is obtained from the Chern--Simons connections through 
\eq{
g_{\mu\nu} = \frac{\ell^2}{2}\,\llangle(A^+ - A^-)_\mu(A^+ - A^-)_\nu\rrangle\,.
}{eq:g}
The (variations of the) Regge--Teitelboim boundary charges are given by
\eq{
\delta Q^{{\pm}}[\epsilon^\pm] ={\pm} \frac{k}{2\pi}\,\oint\llangle\epsilon^\pm\delta A^\pm\rrangle
,}{eq:Q}
where $\epsilon^\pm$ is the transformation parameter whose associated canonical boundary charge is calculated by integrating \eqref{eq:Q}, and $\delta A^\pm$ is in the field space of variations  allowed by the boundary conditions. The integral is over the angular cycle, which we coordinatize by $\vp\sim\vp+2\pi$. The action of the symmetry on any (sufficiently smooth) function $F$ on the phase space is given by the Dirac bracket $\delta_{\epsilon}F=\{Q[\epsilon],F\}$. 

Defining the 1-forms
\be\label{cal-Apm}
{\cal A}_\pm=\zeta^\pm \extd t\pm J^\pm \extd\vp,
\ee
allows to express the near horizon boundary conditions of \cite{Afshar:2016wfy} succintly as 
\eq{
A^\pm = b^{-1}_\pm \,(\extd + \mathfrak{a}_\pm)\, b_\pm,\qquad \mathfrak{a}_\pm=2{\cal A}_\pm\boldsymbol{L_0},
}{eq:bc}
 with some conveniently chosen $SL(2; \mathbb{R})$ group elements $b_\pm$ that depend on the radial coordinate $r$ and are not allowed to vary, $\delta b_\pm =0$. The quantity $\boldsymbol{L_0}$ is the Cartan subalgebra generator of $sl(2; \mathbb{R})$. (Also in the body of the paper we denote generators of various algebras with bold-faced symbols.) The time component of the 1-form ${\cal A}_\pm$ is fixed as well, $\delta\zeta^\pm=0$, so that only its angular component contributes to the charges \eqref{eq:Q}.  These charges are then essentially given by $\vp$-integrals over the functions $J^\pm$, whose zero modes enter in the entropy formula \eqref{eq:intro1}. 
Our normalization of these zero mode charges, $J_0^\pm$, differs from the one used in the entropy formula \eqref{eq:intro1} by the Chern--Simons level, $J_0^\pm={\tt J}_0^\pm/k$. Other than that our conventions are the same as in our previous work \cite{Afshar:2016uax, Afshar:2016wfy, Afshar:2016kjj}.

\section{Locally AdS$_3$ black holes in canonical and microcanonical descriptions}\label{constant-rep}

Our main focus is on constructing semi-classical microstates of Ba\~nados--Teitelboim--Zanelli (BTZ) black holes \cite{Banados:1992wn,Banados:1992gq}. 
To this end we first note that all locally AdS$_3$ geometries form a phase space. A subset of them on which we focus in this work, is associated with unitary Virasoro coadjoint orbits \cite{Balog:1997zz} and contains global AdS$_3$, BTZ \cite{Banados:1992wn,Banados:1992gq} and conic spaces, as well as their conformal descendants \cite{Sheikh-Jabbari:2016lzm}. These correspond to constant representative solutions in the Ba\~nados family of solutions \cite{Sheikh-Jabbari:2016unm,Compere:2015knw}. We also restrict ourselves to cases where both left and right sectors of the solution are in the same family. Within constant representative families one may still consider geometries whose left and right sectors are in different categories. For some background information on Ba\~nados geometries, the Virasoro Hilbert space and Virasoro coadjoint orbits see \cite{Sheikh-Jabbari:2016unm} and appendix \ref{app:A}.

To represent these spacetimes we conveniently use Gaussian normal coordinates \cite{Afshar:2016kjj}
\begin{align}
    \label{afshar2}
 \extd s^2=\extd r^2&-\left[(\ell^2 a^2-\Omega^2)\cosh^2\frac{r}{\ell}-\ell^2a^2\right]\extd t^2 +2\left[\gamma \Omega \cosh^2\frac{r}{\ell} + \ell^2a\omega \sinh^2\frac{r}{\ell}\right]\extd t\extd\varphi \nn\\ &+ \left[\gamma^2\cosh^2\frac{r}{\ell}-\ell^2\omega^2 \sinh^2\frac{r}{\ell}\right]\extd\varphi^2,
\end{align}
where $\varphi\sim\vp+2\pi$ and $\ell$ is the AdS$_3$ radius. Here $a, \gamma, \omega$ and $\Omega$ are functions of $t$ and $\varphi$.  The above metric becomes a solution to Einstein field equations if the near horizon holographic Ward identities\footnote{%
The expression ``near horizon holographic Ward identities'' refers to the fact that the first equality \eqref{EEOM} implies conservation of the canonical charges. Thus, the near horizon conformal Ward identities are analogous to the AdS$_3$ statement that $\partial_\mp T_{\pm\pm}=0$ as a consequence of certain on-shell conditions, often referred to as ``holographic Ward identities'' \cite{Banados:2004nr}.} (relating these functions) hold,
\be\label{EEOM}
\dot{J}^\pm =\pm \zeta'^\pm\qquad\textrm{with}\qquad 2\zeta^\pm\equiv-a\pm\frac{\Omega}{\ell}\quad\textrm{and}\quad 2{J}^\pm\equiv  \frac{\gamma}{\ell}\pm\omega\,,
\ee
where $dot$ and $prime$  denote derivative with respect to $t$ and $\varphi$, respectively. For real functions $a,\omega, \Omega$ and $\gamma$, the metric \eqref{afshar2} represents a black hole with a Killing horizon at $r=0$ which is a BTZ black hole (for constant parameters) or a Virasoro descendant thereof. If these functions are purely imaginary then we obtain instead conic spaces and global AdS$_3$ (for the special choices $-a\ell=\pm i=\gamma\ell^{-1}$, $\Omega=0=\omega$, upon replacing the cycles $t\leftrightarrow\vp$).

In the next two subsections we provide two descriptions of the state-space, one in which the quantities $\zeta^\pm$ are fixed (`canonical') and one where {$\ell\zeta^\pm=J^\pm$ and $J^\pm_0= \frac{1}{2\pi}\int_0^{2\pi} J^\pm$ kept fixed} (`microcanonical'). This thermodynamical nomenclature is justified, since the sum $\zeta^+ + \zeta^-\propto a$ is proportional to the (Unruh-)temperature
\eq{
T_{\textrm{\tiny U}}=\frac{a}{2\pi}
}{eq:TU}
and, as we will discuss momentarily, the canonical boundary charges like the energy and angular momentum of the black hole are given in terms of $J_0^\pm$.

It is worth stressing that boundary conditions based on \eqref{afshar2} with either $a,\Omega$ fixed or $\gamma,\omega$ fixed differ from the usual Brown--Henneaux boundary conditions \cite{Brown:1986nw}. 
Particularly the canonical description of \eqref{afshar2} is very natural from a near horizon perspective and corresponds to the boundary conditions introduced in \cite{Afshar:2016wfy}. To see this explicitly, consider the expansion of the line-element \eqref{afshar2} for $\Omega=0$ (corresponding to a co-rotating frame) around $r=0$,
\eq{
\extd s^2\big|_{\Omega=0} = \extd r^2-(ar)^2\,\extd t^2 + \gamma^2\, \extd\varphi^2 + {\cal O}(r^2).
}{eq:Rindler}
For real $a$ and $\gamma$ the leading order metric \eqref{eq:Rindler} is Rindler space, the universal near horizon approximation to any non-extremal horizon. In the canonical description the Rindler acceleration $a$ is fixed while the horizon area determined by $\gamma$ is allowed to fluctuate. Another property worth mentioning is that the AdS radius $\ell$ drops out of the near horizon line-element \eqref{eq:Rindler}. The properties above provide geometric reasons to refer to the boundary conditions of \cite{Afshar:2016wfy} as `near horizon boundary conditions'.

\subsection{Canonical description}\label{canonical-subsection}

If we take $r=\ell\ln\frac{w}{\ell}$ the metric \eqref{afshar2} can be written as
\be\label{afsharnew}
\extd s^2=\frac{\ell^2\extd w^2}{w^2}-\Big({w {\cal A}_+}-\frac{\ell^2{\cal A}_-}{w}\Big)\Big({w {\cal A}_-} -\frac{\ell^2{\cal A}_+}{w}\Big)\,,
\ee
where we have used the one-forms ${\cal A}_\pm$ defined in \eqref{cal-Apm}. 
The Einstein equations (or near horizon holographic Ward identities) \eqref{EEOM} then reduce to $\extd{\cal A}_\pm=0$. 

A general class of solutions discussed in \cite{Afshar:2016wfy,Afshar:2016kjj}  is given by
\be\label{Afshar-et-al-gauge}
\zeta^\pm={\rm constant \;and\; fixed},\;\; J^\pm=J^\pm(\varphi)\,,
\ee
where $J^\pm(\varphi)$ are periodic functions {allowed to vary}.
This class of solutions corresponds to fixing both the left and right temperatures $T_\pm=\zeta^\pm$. For this reason we call this setup canonical description.\footnote{We note that here the left and right temperatures $T_\pm$ are defined with respect to the near horizon observer \eqref{eq:Rindler}. However, the standard black hole temperatures (the one which is attributed to the presumed asymptotic dual CFT$_2$) is defined by conventions in which the  asymptotic metric is fixed to the so-called asymptotic static frame, that is requiring metric at large $w$ to be $\extd s^2=\ell^2\frac{\extd w^2}{w^2}-w^2 \alpha^2 \extd x^+\extd x^-$ with $x^\pm=t/\ell\pm\varphi,\ x^\pm\in[0,2\pi]$ and  $\alpha$ being  a constant. Then demanding the asymptotic and near horizon temperatures to match, which is the convenient choice for discussing black hole thermodynamics,  we need to relate $\zeta^\pm$  to $J^\pm$ as $\ell\zeta^\pm=J_0^\pm\equiv \frac{1}{2\pi}\int\limits_0^{2\pi} \extd\varphi\ J^\pm({\varphi})$. 
}
For constant, real $J^\pm(\varphi)$ the above geometries are BTZ black holes, with  mass and angular momentum given by
\be
M=\frac{1}{4G}\,\big((J_0^+)^{2}+(J_0^-)^{2}\big),\qquad J=\frac{\ell}{4G}\,\big((J_0^+)^{2}-(J_0^-)^{2}\big)\;.
\ee

The choice $(J_{{0}}^{\pm})^2=-\tfrac14$ yields global AdS$_3$, while for constant $J^\pm$ restricted to $-\tfrac14<(J^\pm)^2<0$ we have spaces with conic deficits. 
\subsection{Microcanonical description}\label{microcanonical-subsection}

In the subsection above we presented the geometries in a canonical description [keeping fixed the Unruh-temperature \eqref{eq:TU}]. For the microcanonical description it is useful to rewrite the above class of solutions in a slightly different way. {In contrast to \eqref{Afshar-et-al-gauge}, here we consider} the following solution to $\extd{\cal A}_\pm=0$ on the cylinder:
\be\label{sol-mic-const}
\ell\zeta^\pm= J^\pm=J^\pm(x^\pm)\,,\qquad{\cal A}_\pm= J^\pm \extd x^\pm\,,
\ee
where $x^\pm\in[0,2\pi]$ are light-cone coordinates on the cylinder. The zero mode solution $\ell\zeta^\pm=J_0^\pm$ is both in the constant (and fixed) chemical potential family \eqref{Afshar-et-al-gauge} and also in the Ba\~nados family \eqref{generic-Banados-geometry}. This is easy to see by replacing the radial coordinate $w$ in \eqref{afsharnew} by $z$ through $z^2=w^2J_0^+J_0^-$ which yields
\be\label{TBTZ}
\extd s^2=\frac{\ell^2\extd z^2}{z^2}-\left({z\extd x^+} -\frac{\ell^2{L_0^-}\extd x^-}{z}  \right)\left( {z\extd x^-} -\frac{\ell^2{L_0^+} \extd x^+}{z}  \right)\,,
\ee
where $L_0^\pm=J_0^{\pm\,2}$. 
As discussed, depending on the value of $L_0^\pm$ one obtains three classes of geometries: $L_0^\pm=-\tfrac{1}{4}$,  $L_0^\pm \in (-\tfrac{1}{4},0)$ and $L_0^\pm \geq 0$, respectively, correspond to global AdS$_3$, conic deficits and BTZ black holes. For easier reference and comparison between different variables we collect the various conditions on free functions (for zero-mode solutions) in table \ref{tab:1}. {The most general solution obeying \eqref{sol-mic-const} is related to Ba\~nados geometries \eqref{generic-Banados-geometry} upon a cumbersome coordinate transformation. The result is relating the functions $L^\pm$ and $J^\pm$  through a twisted Sugawara construction \cite{Sheikh-Jabbari:2016unm, Afshar:2016wfy, Afshar:2016kjj}\,,
\bea\label{J-L}
L^\pm=J^{\pm\,2}+ J^{\pm\,\prime}\,.
\eea
These most general solutions constitute all Virasoro descendants of the three class of constant $J^\pm$ geometries discussed above. }
\begin{table}
\begin{tabular}{|l||c|c||c|c||c||}\hline
        & $a$ & $\gamma$ & $\zeta^\pm$ &$J^\pm$ & $L^\pm$ \\\hline
global AdS$_3$ & $\mp i\ell^{-1}$& $\pm i\ell$ & $\ell\zeta^+=\ell\zeta^-=\pm \tfrac {i}{2}$ & $J^+ = J^-=\pm \tfrac i2$ & $L^+ = L^- =-\tfrac14$ \\
Conic   & imaginary & imaginary & imaginary, $|\zeta^\pm|<\tfrac12$ & imaginary, $|J^\pm|<\tfrac12$ & $-\tfrac14<L^\pm<0$\\
BTZ     & real & real & real & real & $L^\pm \geq 0$\\
\hline
\end{tabular}
\caption{Conditions on free functions for zero mode solutions in various formulations; note our normalization $J^\pm={\tt J}^\pm/k$ as compared to the entropy formula \eqref{eq:intro1}.}
\label{tab:1}
\end{table}

As another way to find the most general solution in this class
we may recall the Poincar\'e Lemma and the set of solutions \eqref{sol-mic-const}. Then, the solutions on the cylinder to $\extd{\cal A}=0$ is of the form
\be\label{Sol-mic2}
{\cal A}_\pm= J_0^\pm \extd h_\pm (x^\pm), \qquad \textrm{with} \qquad h_\pm(x^\pm+2\pi)=h_\pm(x^\pm)+2\pi\,. 
\ee
The above result has a very simple physical interpretation: the solution \eqref{sol-mic-const} can be generated by an orientation preserving  conformal transformation on the cylinder,
\be\label{confmap}
x^\pm\mapsto h_\pm(x^\pm)\,,\qquad h'_\pm>0,
\ee
where \textit{prime} denotes derivative with respect to the argument. 
Using the finite conformal transformation of the functions $L^\pm(x^\pm)$ appearing in the Ba\~nados form \eqref{generic-Banados-geometry} reviewed in appendix \ref{app:A}, we have
\begin{align}\label{L-h-J0}
L^\pm(x^\pm) ={J_0^\pm}^2 {h_\pm'}^2 -\frac{1}{2} \left[\frac{h_\pm'''}{h'_\pm}-\frac{3}{2}\left(\frac{h_\pm''}{h_\pm'}\right)^2\right]\,.
\end{align}
This is the most general form of functions $L^\pm$ in  constant representative Virasoro coadjoint orbits \cite{Afshar:2015wjm, Sheikh-Jabbari:2016unm,Balog:1997zz}. 
Correspondingly, using \eqref{J-L} the  $J^\pm$-fields are  obtained in terms of $h_\pm$ in a conformal family (orbit):
\be\label{J-h-Phi0}
J^\pm(x^\pm)=\pm J_0^\pm h_\pm'-\frac12 \frac{h_\pm''}{h_\pm'}.
\ee
Note that the above $J^\pm\big(x^\pm\big)$ is not the same function $J^\pm(\varphi)$ defined in the previous subsection in the canonical description. In the canonical description the left and right temperatures $\zeta^\pm$ are constant (held fixed), whereas here we fix corresponding energies $J_0^\pm$. That is why we refer to this choice as microcanonical descripition. To distinguish the microcanonical and canonical descriptions, we denote the canonical degrees of freedom by the functions $J^\pm(\vp)$ (as we did in subsection \ref{canonical-subsection}) and the microcanonical ones by $h_\pm$, as defined in \eqref{Sol-mic2} and \eqref{J-h-Phi0}.

\subsection{Canonical to microcanonical map}\label{sec:can-to-mcan}

In \cite{Afshar:2016wfy, Afshar:2016kjj}, it has been shown that the formula \eqref{J-L} holds independently of the choice of our boundary conditions in fixing or varying variables. The $L^\pm$, which may be viewed as the energy momentum tensor of a CFT$_2$ on the cylinder, admit canonical or microcanonical descriptions. The canonical description is appropriately governed by the $J^\pm=J^\pm(\varphi)$  \eqref{Afshar-et-al-gauge} as its dynamical variable (field) and the microcanonical description by the $h_\pm(x^\pm)$ field given in \eqref{J-h-Phi0}. While the two fields $h_\pm(x^\pm)$ and $J^\pm(\varphi)$ are different, we expect these two to be describing the same physical system {at constant time slices}.  
In this part we discuss how these two descriptions are related to each other. For notational simplicity, let us suppress the $\pm$ index denoting the left-right sectors.
Equating $J_{\textrm{\tiny can}}$ and $J_{\textrm{\tiny mic}}$ we get the canonical to microcanonical map, which relates the corresponding fields at constant time slices:
\be\label{J-h-Phi}
J(\varphi)=\pm J_0 h'-\frac12 \frac{h''}{h'}\,,\qquad\text{at constant time slice.}
\ee
Note that, as the above also shows, there are two physically equivalent such $J(\phi)$ for a given function $h(\phi)$. These two solutions are related through $J_0\to -J_0$. As we will discuss in the next section there are physically sensible and indeed very relevant set of solutions corresponding to imaginary $J_0$. In such cases the two solutions are complex-conjugate of each other.

As it will become clear in section \ref{sec:more-on-HCG}, it is convenient to define a new field $\Phi$ as integral of $J(\phi)$:
\be\label{Phi-field}
\Phi\equiv\int^\phi J= \Phi_0 \pm J_0 h-\frac12 \ln h'\,.
\ee
Note that since $h'>0$ the log-term is always real-valued and well-defined; moreover,
\be\label{Phi-periodicity}
\Phi(\phi+2\pi)=\Phi(\phi)\pm 2\pi J_0\,.
\ee
See also \cite{NavarroSalas:1999sr} for a similar discussion.

\section{Symplectic symmetry and phase space for canonical description}\label{sec-canonical-Hilbert}

In this section we focus on black holes in the canonical description introduced in section \ref{canonical-subsection}. This set of solutions may be viewed as a phase space \cite{Afshar:2016wfy, Sheikh-Jabbari:2016unm}. We discuss the corresponding symplectic symmetries and the Hilbert space associated with the symmetry algebra. Since all our formulas apply in the same way in the plus and minus sectors, from now on we drop the $\pm$ decorations and treat both chiral sectors simultaneously and on equal footing.
\begin{figure}
    \centering
    \includegraphics[width=9cm]{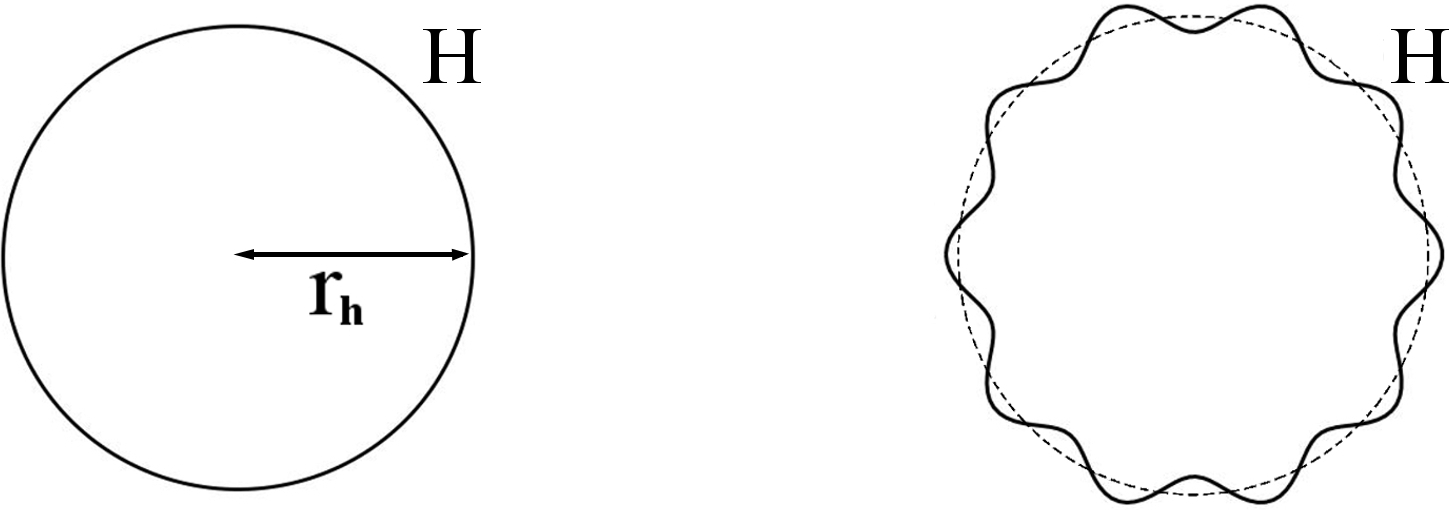}
    \caption{BTZ horizon without (left) and with (right) soft hair excitations}
    \label{fig:1}
\end{figure}
The left figure \ref{fig:1} depicts a BTZ black hole (with horizon radius $r_h$) and the right one a generic descendant of the same black hole. The wavy line shows the horizon described by $r=r(\varphi)$ curve in the coordinate system used in \eqref{afshar2} or \eqref{generic-Banados-geometry}. For the explicit equation for the horizon curve see \cite{Sheikh-Jabbari:2016unm}. In other words, a generic state in $\cH_{\textrm{\tiny BTZ}}$ may be depicted through the figure in the right.

\subsection{Canonical symplectic symmetry algebra}\label{sec:can-sympl}

It was shown in \cite{Afshar:2016wfy} that the algebra of symplectic symmetries\footnote{In three dimensions there is a simple way to make the boundary charges independent of the radial coordinate. In that case we refer to the asymptotic symmetries as ``symplectic symmetries'' \cite{Compere:2015knw,Compere:2015mza,Compere:2015bca,Compere:2014cna}, since they are a property of the physical phase space, which may be thought to be localized at any hypersurface of constant radius, not just at the asymptotic boundary. In this context ``boundary gravitons'' may also be called ``holographic gravitons'' \cite{Compere:2015knw}. So the usual meme that ``the dual field theory lives at the asymptotic boundary'' could be replaced by the statement ``the dual field theory lives at the horizon'', without changing any of the formulas. Which of these interpretations (if any) is more suitable depends on the precise context and the questions one is asking. For some of the questions we are addressing in the present work the near horizon interpretation seems more to the point.
} associated with geometries given by \eqref{afsharnew}-\eqref{Afshar-et-al-gauge} consists of two commuting $\hat u(1)_k$ current algebras, with level $k=c/6$,
\be\label{J-algebra}
[\bJ_n,\,\bJ_m]=\frac{c}{12} \,n\, \delta_{n,-m}\,,
\ee
and $c$  is the Brown--Henneaux central charge \eqref{J-c-intro}.
Geometries with functions $J({\varphi})$ correspond to the states in the Hilbert space of this algebra such that expectation values of $\bJ_n$ in terms of Fourier modes $J_n$ of the charges $J(\varphi)$ are given by
\be\label{J-field-opt-algebra}
\langle \bJ_n\rangle = \frac{c}{6} J_n\qquad\textrm{with}\quad 
J({\varphi}) =\sum_{n\in\mathbb{Z}} J_n e^{in{\varphi}}\,.
\ee
In accordance with  \eqref{J-algebra} and \eqref{J-field-opt-algebra}, one obtains the Poisson bracket between these variables on the classical phase space $J(\vp)$,
\be\label{J-phi-algebra}
\{J({\varphi_1}), J({\varphi_2})\}=\frac{6\pi}{c} \cdot   \partial_{\varphi_2}\delta({\varphi_1}-{\varphi_2})
\ee

In a similar way one can quantize $L(\vp)$ or its Fourier modes, using \eqref{J-L}:
\be\label{twisted-sugawara}
\bL_n\equiv \frac6{c}\sum_{p\in\mathbb{Z}} \!\bJ_{n-p}\bJ_p + in\bJ_n
\ee
which are also conserved and satisfy the algebra,
\be\label{L-J-algebra}
[\bL_n,\,\bL_m]=(n-m)\bL_{n+m}+\frac{c}{12}\,n^3\,\delta_{n,-m}\qquad
[\bL_n,\,\bJ_m]=-m\,\bJ_{n+m}+ i\frac{c}{12} \,m^2\, \delta_{n,-m}\,.
\ee

Note also that $\bJ_0$ is a central element of the algebra, i.e., it commutes with all $\bJ_n$ and $\bL_n$.  In the language of representation theory, each Virasoro coadjoint orbit is generically specified by a real continuous number $J_0$ \cite{Sheikh-Jabbari:2016unm,Balog:1997zz, Compere:2015knw}.

Reality and smoothness of the black hole metrics in  \eqref{afshar2} requires $J({\varphi})$ to be in general a real periodic function. As discussed in the previous section, the set of geometries in \eqref{afshar2} can also describe spaces with conic singularities if we allow for ${J_0}^2$ to take negative values. However, unitarity of the corresponding Virasoro coadjoint orbits, i.e. requiring that the value of $L(\vp)$ is bounded from below in the orbit, restricts its value as ${J_0}^2\geq -\tfrac14$ corresponding to conic spaces and global AdS.
Moreover, reality of  other modes in $J$ functions, i.e. $J_n$ for $n\neq 0$, requires $J_{-n}={J_n}^\ast$. For the associated charge operators $\bJ_n$, we then learn that the zero mode generators $\bJ_0$ can be hermitian or anti-hermitian, with eigenvalues limited by unitarity, and  the remaining generators must obey $(\bJ_n)^\dagger=\bJ_{-n}$ for $n\neq 0$ (see appendix \ref{appendix-D} for more discussions). Since the commutator of two $\bJ_n$ currents is proportional to $\frac{c}{12}$ --- see eq.~\eqref{J-algebra}, eigenvalues of the `number operator',
\eq{
\bN=\sum_{n>0}\bJ_n^\dagger \bJ_n\,,
}{eq:NumOp} 
take values in integer-multiples of $\frac{c}{12}$. For integer $\hat u(1)_k$ levels $k$, the Brown--Henneaux relation \eqref{J-c-intro} implies that these eigenvalues are either half-integer or integer. We shall use the properties summarized above in our proposed correspondence between two different weakly coupled descriptions in section \ref{sec:BH-CG-duality} below.
\begin{figure}
    \centering
    \includegraphics[width=9cm]{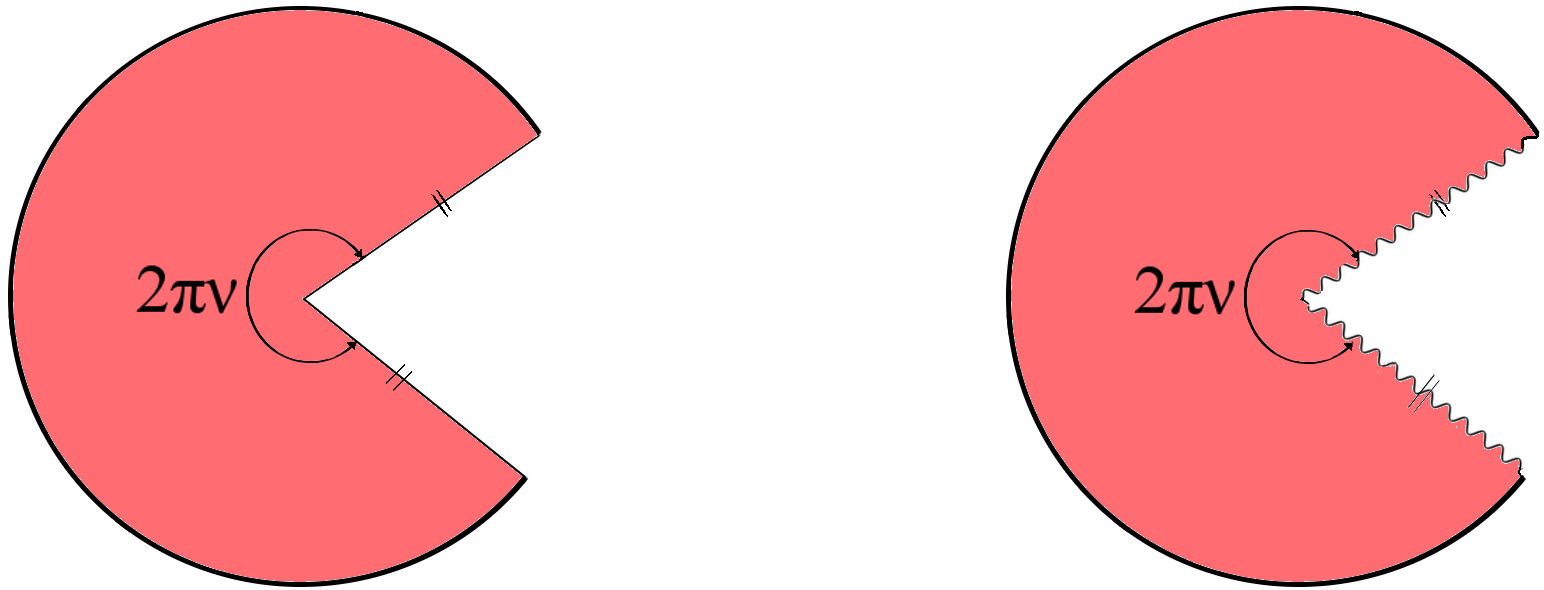}
    \caption{Particle on AdS$_3$ (left) and a Virasoro descendant thereof (right)}
    \label{fig:2}
\end{figure}

Figures \ref{fig:2} show static conic spaces with deficit angle $2\pi(1-\nu)$. The left figure shows a particle on AdS$_3$ and the right figure its (Virasoro) descendant. In the above the circle is the Poincar\'e disk which is the constant time slice on global AdS$_3$ and the conic space is obtained through identification of the edges of the angle $2\pi(1-\nu)$. In other words, the left figure with all possible values for $\nu\in(0,1]$ may be viewed as a generic state in the Hilbert space $\cH_{\textrm{\tiny CG}}$.

\subsection{Hilbert space of geometries in canonical description}\label{sec:3-2}

As the Ba\~nados family of geometries realizes the representation of two copies of the Virasoro algebra --- see appendix \ref{app:A} for a review --- the non-extremal family of BTZ geometries introduced in \eqref{Afshar-et-al-gauge} should fall into the representations of the algebra \eqref{J-algebra} and \eqref{L-J-algebra}. Here, we review the construction of this representation as it will be essential for the precise statement of the horizon fluff proposal. 

Dealing with a simple algebra of creation-annihilation operators \eqref{J-algebra}, it is straightforward to construct its representations and the corresponding Hilbert space, which  in a convenient notation may be denoted as $\cH_{\bJ}^+\otimes \cH_{\bJ}^-$. Again concentrating on one sector, $\cH_{\bJ}$, following \cite{Afshar:2016uax, Sheikh-Jabbari:2016npa}, we first define the vacuum states $|0; J_0\rangle$ as specific highest weight states
\be\label{Asymp-vacuum-def}
\bJ_n |0; J_0\rangle=0\qquad  \forall n>0, \qquad \textrm{and} \qquad \bJ_0|0; J_0\rangle=\frac{c}{6}J_0|0; J_0\rangle\,,
\ee
with the normalization $\langle 0; J'_0|0; J_0\rangle=\delta_{J'_0,J_0}$. A generic state $|\{n_i\}; J_0\rangle$ is produced by acting with any number of creation operators $\bJ_{-n_i}$ ($n_i>0$) on this vacuum:
\be\label{generic-NH-state}
 |\{n_i\}; J_0\rangle =  {\cal N}\prod_{\{n_i>0\}} \bJ_{-n_i} |0; J_0\rangle \qquad \forall |\{n_i\}; J_0\rangle \in {\cal H}_{\bJ}\,,
\ee
where ${\cal N}$ is a normalization factor fixed such that
\be
\langle \{{n'}_i\}; J'_0|\{n_i\}; {J}_0\rangle=\delta_{\{{n}'_i\},\{{n}_i\}}\delta_{J'_0,J_0},
\ee
and $\delta_{\{{n}'_i\},\{{n}_i\}}$ is the product of all the Kronecker $\delta_{n'_i,n_i}$.

For the BTZ black hole family $J_0$ is a real number (see table \ref{tab:1} on p.~\pageref{tab:1}), which may be conveniently chosen to be positive. States or geometries associated with $J_0$ and $-J_0$ are related to each other through parity and are not physically distinct. From table \ref{tab:1} we see that $J_0$ can take imaginary values too, $J_0=i\nu/2$. Unitarity of the corresponding Virasoro representation restricts $-1\leq\nu\leq 1$, which covers precisely the cases displayed in table \ref{tab:1}. One can conveniently restrict $\nu$ to non-negative values, $0\leq \nu \leq 1$. Saturation corresponds to global AdS$_3$ for $\nu=1$, and to massless BTZ for $\nu=0$. The remaining cases, parametrized by the spectral flow parameter $0<\nu<1$ correspond to spaces with conic deficits \cite{Maldacena:2000dr}, which can be interpreted as massive particles on AdS$_3$.

\begin{figure}
    \centering
    \includegraphics[width=14cm]{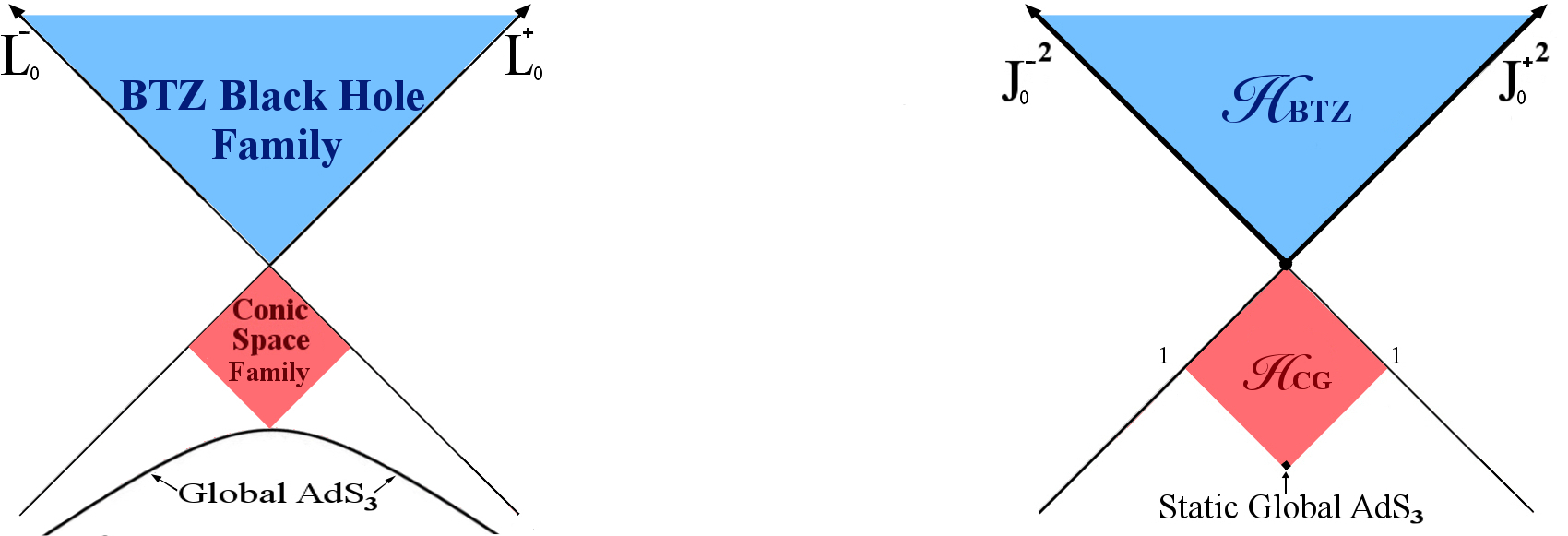}
    \caption{Hilbert spaces $\cH_{\textrm{\tiny Vir}}$ (left) and $\cH_{\bJ}$ (right)}
    \label{Hvir-L-J}
\end{figure}
The left (right) figure \ref{Hvir-L-J} sketches the Hilbert spaces $\cH_{\textrm{\tiny Vir}}$ ($\cH_{\bJ}$). Each point in the colored region of the right figure corresponds to various states in the Ba\~nados family of black holes (see appendix \ref{app:A}). These states include usual BTZ black holes and their Virasoro descendants (the blue region) and the conic space and global AdS$_3$ and their Virasoro descendants (the red region). Each point in the colored region of the right figure shows a Virasoro coadjoint orbit specified with values $J_0^\pm$. We note that while each point in the right figure shows a Virasoro orbit, each point in the left figure corresponds to states in different Virasoro orbits with the same values for $L_0^\pm$.

To summarize, all the unitary states in the Hilbert space of the canonical algebra $\cH_{\bJ}$ belong either to the black hole subspace $\cH_{\textrm{\tiny BTZ}}$ or the remainder $\cH_{\textrm{\tiny CG}}$, i.e.,
\be
\cH_{\bJ}=\cH_{\textrm{\tiny BTZ}}\cup \cH_{\textrm{\tiny CG}},\qquad \cH_{\textrm{\tiny CG}}=\cH_{\textrm{\tiny Conic}}\cup\cH_{\textrm{\tiny gAdS}},
\ee
where $\cH_{\textrm{\tiny BTZ}}$ consists of all states with real-positive  $J_0$, corresponding to BTZ black holes and their conformal descendants, and $\cH_{\textrm{\tiny CG}}$  with  imaginary $J_0=i\nu/2,\ \nu\in(0,1]$; the $\nu\in (0,1)$ fall into $\cH_{\textrm{\tiny Conic}}$ corresponding to conic spaces and their Virasoro descendants and  $\nu=1$ case corresponds to global AdS$_3$ and its descendants falling into $\cH_{\textrm{\tiny gAdS}}$. We note that the $\cH_{\bJ}$, and its three categories, $\cH_{\textrm{\tiny BTZ}},\ \cH_{\textrm{\tiny Conic}},\ \cH_{\textrm{\tiny gAdS}}$ may also be viewed as Virasoro coadjoint orbits (\emph{cf}. discussions in appendix \ref{Vir-coadj-orbits} or in \cite{Sheikh-Jabbari:2016unm}).  More discussions on the geometry of conic defects may be found in appendix \ref{app:Conic}. 

In figure \ref{Hvir-L-J} we have depicted the $\cH_{\textrm{\tiny BTZ}}$ and $\cH_{\textrm{\tiny CG}}$ in two different bases, in $L$-basis and in $J$-basis. {Note that each point in the right figure corresponds to a Virasoro coadjoint orbit with a given $J_0^\pm$ while a generic point in the left figure corresponds to geometries which have the same $L_0^\pm$ and hence fall in different Virasoro orbits.  The points on the positive $(J_0^\pm)^2$ axis, which have vanishing $J_0^\mp$ and has been depicted by a thicker line becauase they may correspond to both extremal BTZ family or self-dual AdS$_3$ orbifold. (See \cite{Sheikh-Jabbari:2016npa, Sheikh-Jabbari:2016unm} for more discussions).} We also note that, as discussed in \cite{Sheikh-Jabbari:2016npa},  states with $J(\phi)=\frac{6}{c}\langle \bJ(\phi)\rangle$ are \emph{coherent states} in $\cH_{\bJ}$, as they are states which diagonalize $\bJ_n$ and are its eigenvectors.

\section{Conic spaces, corresponding coherent states, orbits  and Hilbert space}\label{sec:more-on-HCG}

Here we focus more on the $\cH_{\textrm{\tiny CG}}$ subspace  of the full Hilbert space $\cH_{\bJ}$. Unlike the black hole family, these states correspond to anti-hermitian $\bJ_0$ and imaginary $J_0$. It is hence desirable to seek a 
field representing this family, different than $J(\phi)$ or $\bJ(\phi)$.\footnote{In our conventions operator-valued fields are defined as $\bJ(\phi)=\sum_n \bJ_n e^{in\phi}$ such that 
$\langle \bJ(\phi)\rangle=\frac{c}{6}J(\phi)$.\label{operator-J}}

\subsection{Wilson lines as primary  fields of weight one with twisted boundary conditions} 
Given the energy momentum tensor $L(\phi)=J'(\phi)+J^2(\phi)$, the fields $J(\phi)$  are quasi-primary. This is seen from commutation relations \eqref{L-J-algebra}
 or more directly by using \eqref{J-phi-algebra} to find that under a conformal transformation\footnote{In the semi-classical case of large $J$, when the anomalous term $\epsilon''$ in $\delta_\epsilon J$ may be ignored, the quantity $J$ also becomes a primary field of weight one and provides an appropriate description for the black hole family.}
\be\label{delta-J}
\delta_\epsilon J(\phi)=(\epsilon J)'-\epsilon''/2\,,
\ee
where we used the fact that $\delta_{\epsilon}F=\{Q[\epsilon],F\}$.
One can instead construct a new field which is  a \emph{primary of weight one};
\be\label{W-nu-def}
{\cal W}\equiv e^{-2\Phi}\,,\quad\qquad\Phi=\int^\phi J=\Phi_0+J_0\phi-i\sum_{n\neq0}\frac{J_n}{n}e^{in\phi}.
\ee
To see that $\cW$ is indeed a  primary of weight one, we use \eqref{delta-J} to derive $\delta_\epsilon (\int^\phi J)=\epsilon J-\epsilon'/2$, yielding 
\be
\label{eq:primary}
\delta_\epsilon\cW=(\epsilon \cW)'\,. 
\ee
Geometrically, cases with $J_0$ and $-J_0$ are mapped to each other by a $\mathbb{Z}_2$-symmetry ($\phi \to -\phi$) and are hence not physically distinct. 
That is why we can label the $\cW^\pm$ fields with the index $\pm$; here we abbreviate $\cW^\pm\equiv\cW(\phi;\pm\Phi_0,\pm J_0)$. However, by imposing an appropriate condition we relate and identify $\pm$ cases. From \eqref{W-nu-def} we learn that $\cW^\pm$ fields have the following periodicity and hermiticity properties 
\be\label{W-periodicity}
\cW^{\pm }(\phi+2\pi)=e^{\mp 4\pi J_0}\cW^{\pm }(\phi)\,,\qquad (\cW^{\pm })^\ast=\cW^{\mp }\,.
\ee
The periodicity property in \eqref{W-periodicity} shows that while one can define $\cW^\pm$ fields for real $J_0$ too, they are particularly suitable fields for imaginary $J_0$; namely $J_0=i\nu/2$  corresponds to geometries without horizon but having $2\pi(1-\nu)$ conical defects for  $\nu\in(0,1)$. Since the primary fields $\cW^{\pm}$ have twisted boundary conditions \eqref{W-periodicity} we refer to them as ``twisted primaries''.\footnote{In general, $\Psi_b=e^{-b\int^\phi J}$ is a primary field of weight $b/2$. In particular the case of $b=1$ leads to a fermionic primary field with the periodicity $\Psi(\phi+2\pi)=e^{\mp i\pi  \nu }\Psi(\phi)$. Such fermionic fields are anti-periodic for global AdS$_3$ ($\nu=\pm 1$) while for the massless BTZ ($\nu=0$) they are periodic.  One may hence label the global AdS$_3$ and the massless BTZ states as NS and R-vacuum, respectively \cite{Maldacena:2000dr, David:2002wn}.\label{footnote:Fermion-primaries} } The notion of conic deficit is directly related to the twisted periodicity \eqref{W-periodicity}. Similar fields have also been introduced and discussed in \cite{NavarroSalas:1999sr}.

\subsection{Quantization of the $\cW$ fields}\label{Quant-W-section}

As noted above, the $\cW$-fields under conformal transformation generated by $Q[\epsilon]=\frac{k}{2\pi}\oint \epsilon L$ are primary fields of unit weight. One may examine how they transform under $Q[\eta]=\frac{k}{2\pi}\oint \eta J$ generated by the $J$-fields. Since we have \cite{Afshar:2016wfy, Afshar:2016kjj}
\be
\label{eq:deJ}
\delta_\eta J={+}\eta'/2,
\ee
we get,
\be\label{delta_etaW}
\delta_{\eta_\pm} \cW^\pm ={-}\eta_\pm \cW^\pm
\ee
with 
\be
\label{eq:importantenough}
\eta_\pm=\int \eta'\pm i\eta_0\,, 
\ee
where $\pm i\eta_0$ are zero modes of the functions $\eta_\pm$.
The first term on the right hand side of \eqref{eq:importantenough} follows from the transformation of $J$ \eqref{eq:deJ} and the definition of $\cW$ \eqref{W-nu-def}. The second term is less obvious. The imaginary unit as well as the $\pm$ sign come from the fact that the zero mode of $J$ is imaginary and from the  consistency of equation \eqref{delta_etaW} under complex conjugation. Thus, the zero mode of the $\Phi$ field, $\Phi_0$, should be purely imaginary.
We note also in passing that the zero mode of $\eta$ (essentially $\eta_0$) does not appear in an analysis that involves any generic local combination of $J(\phi)$ fields. In other words, $\bJ_0$ commutes with any local operator made out of the field $\bJ(\phi)$. Note however, that $\cW$ is local only in $\Phi$, which in turn is a non-local function of $J(\phi)$, see \eqref{W-nu-def}.

In order to quantize the theory we promote $\cW$ to an operator. This may be done in a consistent way through 
\be\label{bcW}
\bcW^+(\phi)=(\bcW^-(\phi))^\dagger \equiv \colon \!\!e^{-2\mathbf{\Phi}}\!\!\colon
\ee
where $\colon \colon$ denotes the appropriate ordering guaranteeing that $\bcW$ is a primary of weight one. Recalling the definition of coherent states the operator \eqref{bcW} (acting on the vacuum) generates a coherent state in $\cH_{\bJ}$. This coherent state describes a coherent collection of particles on AdS$_3$ and their Virasoro descendants and geometrically corresponds to geometries with a given function $J(\phi)$ (see \cite{Sheikh-Jabbari:2016npa} for more discussions). Below we make this picture more precise by analyzing commutators of $\bcW$ and $\bJ$ fields and by introducing a unique vacuum state.

In this sector $\mathbf{\Phi}_0, \bJ_0$ are both anti-hermitian while the $\bJ_n$ satisfy the algebra \eqref{J-algebra} and\footnote{The operator $\mathbf{\Phi}_0$ moves between the Virasoro orbits, labeled by $J_0$, and thus does not commute with $\bJ_0$. Using $\mathbf{\Phi}_0$ one can construct a `shift operator' of the form $U_\theta\equiv e^{-\theta\mathbf{\Phi}_0}$, where $\theta$ is a continuous real parameter. This operator generates the `spectral flow' map on the algebra \eqref{J-algebra} and \eqref{L-J-algebra}
\begin{equation*}
   \bL_n\to \bL_n-i\theta \bJ_n-\theta^2\frac{c}{6}\delta_{n,0}\qquad
   \bJ_n\to \bJ_n-i\theta\frac{c}{12} \delta_{n,0}\,,
\end{equation*}
which leaves the algebra invariant \cite{Schwimmer:1986mf}. Obviously this map is only unitary on $\cH_{\bJ}$ when $\mathbf{\Phi}_0$ is anti-hermitian, which is the case for the conic space sector where $\bJ_0$ is also anti-hermitian.\label{spectral-flow-footnote}
}
\be\label{phi0-J0}
[\mathbf{\Phi}_0, \bJ_0]=i\frac{c}{12}\,,\qquad \langle\mathbf{\Phi_0}\rangle=\Phi_0\,.
\ee
Note that with our conventions the operators associated to the charges $\bJ_n$ and the corresponding classical (or expectation) values are related by a factor of $c/6$ and hence $\{\Phi_0, J_0\}=1/2$ takes its standard form.
Equation \eqref{delta_etaW} then yields the following quantum commutators
\begin{align}\label{J-W-commutator}
[\bJ_n,\bcW^{\pm}_m]= {-}\,i\bcW^{\pm}_{n+m}\; (\forall n\neq 0)\,,
\qquad [\bJ_0,\bcW^{\pm}_n]=\pm \,i \frac{c}{6}\bcW^{\pm}_{n}\,,
\end{align}
where $\bcW^{\pm}_{n}$ are Fourier modes of the $\bcW^\pm(\phi)$ field.
Therefore, $\bcW^{\pm}_n$ are not in the enveloping algebra of $\bJ_n$ which has $\bJ_0$ as its center. This may be understood recalling the definitions \eqref{W-nu-def} and the commutator \eqref{phi0-J0}. Note in particular that the $[\bJ_0, \bcW]$ commutator is of order $c$ while $[\bJ_n, \bcW], n\neq 0$ is of order one.
Moreover, we learn that 
\be\label{Wn-states-J0}
\bJ_0(\bcW^{\pm}_n|0;J_0\rangle)= \frac{c}{6}(J_0\pm i)\bcW^{\pm}_n|0;J_0\rangle, 
\ee
and hence states $\bcW^{\pm}_n|0;J_0\rangle$ with real $J_0$ do not fall into $\cH_{\bJ}$, as $J_0\pm i$ is a complex number, while states in $\cH_{\bJ}$ have real or imaginary $J_0$. So, $\bcW^{\pm}_n$ is not a well-defined operator in the black hole sector. This latter may also be understood recalling the periodicity relation \eqref{W-periodicity} and that $\cW(\phi)$ in that case receives an exponential scaling every time we shift $\phi$ by $2\pi$. For imaginary $J_0$, with $J_0=i\nu/2$, the eigenvalue equation \eqref{Wn-states-J0} implies that the action of $\bJ_0$ on $\bcW^{\pm}_n$ shifts $\nu$ to $\nu\pm 2$ and hence takes us out of the unitarity range $[-1,1]$. This means that $\bJ_n$ is not a well-defined operator on the Hilbert space of states constructed through $\bcW^{\pm}_n$ and/or $|0;J_0\rangle$ is not a good vacuum state for this Hilbert space. To construct this Hilbert space we hence need a different vacuum state other than $|0;J_0\rangle$. We shall return to this issue in the next subsection.  We should however comment that it makes sense to label $\bcW_n$ by $\pm\nu$,  despite the fact that $\bJ_0$ does not commute with them, because \eqref{Wn-states-J0} implies that $\nu$ is shifted by $2$ units and hence values of $\pm\nu$ in $[-1,1]$ range are still well-defined. Hence, hereafter when there is no confusion, we will view $\bcW^\pm$ in a basis where $\bJ_0$ is diagonal, denoting our fields by $\bcW^{\pm\nu}(\phi)$ and their Fourier modes by $\bcW_n^{\pm\nu}$. We also note that the right equation in \eqref{W-periodicity} yields,
\be\label{W-dagger-W-conjugates}
(\bcW_n^{\pm\nu})^\dagger=\bcW_{-n}^{\mp\nu}\,.
\ee

Next, we need the commutators of the $\bcW_n$ among themselves. To this end, we use the consistency condition for the algebra with $\bJ_n,\ \bcW_n^\pm$ generators, with $[\bJ,\bJ]$ and $[\bJ,\bcW]$ commutators given in \eqref{J-algebra} and \eqref{J-W-commutator} and examine the Jacobi identity involving $\bJ,\bcW,\bcW$. We work in the large $c$ limit and drop subleading terms. In that limit we use the scalings $\bJ_{n\neq0}\sim {\cal O}(1)$, $\bJ_0\sim{\cal O}(c)$, which permit us to ignore the commutation relations of $\bJ_{n\neq0}$ with $\bcW_n$ since
\eq{
[\bJ_n, \bcW^{\pm\nu}_m] = \pm i\frac{c}{6} \bcW^{\pm\nu}_{m}\delta_{n,\,0}+{\cal O}(1)\,.
}{eq:intermediate0}
This allows us to focus on the Jacobi identity involving $\bJ_0, \bcW_n^\nu, \bcW_m^{-\nu}$, which holds provided the commutator between two $\bcW$-fields has the following form
\eq{
[\bcW_n^\nu, \bcW^{-\nu}_m] = c\,A(n,m) + i\, B(n,m)\bJ_0,
}{eq:intermediate1}
with so far arbitrary functions $A(n,m)$, $B(n,m)$. The factor $c$ was pulled out of the central term $A$ to make both terms of them same order in a large $c$ expansion (also, it is physically plausible that the central term scales with the central charge $c$). The factor $i$ was pulled out of $B$ due to anti-hermiticity of $\bJ_0$.
Consistency with hermiticity \eqref{W-dagger-W-conjugates} holds if both these functions are proportional to $\delta_{n,\,-m}$, i.e., $A(n,m)=a(n)\delta_{n,\,-m}$, $B(n,m)=b(n)\delta_{n,\,-m}$, with $a(n)^\dagger=a(n)$ and $b(n)^\dagger=b(n)$. With hindsight, we fix the function $a(n)$ linear in $n$ and $b(n)$ constant, with the following precise values,
\begin{align}
[\bcW_n^{\pm\nu}, \bcW^{\mp\nu}_m]
=\left(\frac{c}{12} n\mp i\bJ_0\right)\ \delta_{n,-m}\,. 
\label{Wnu-algebra1}
\end{align}
Also the Jacobi identities involving only the fields $\bcW$ then hold to leading order in $c$, by which we mean concretely that all terms of ${\cal O}(c^2)$ [but not necessarily ${\cal O}(c)$] cancel in this Jacobi identity.

The fact that the ``interaction terms'' $\langle\cW\cW J\rangle$ and $\langle\cW J J \rangle$ are suppressed by factors of $1/c$ suggests that in this regime $\cW$-fields, in parallel to $J$-fields for the black hole sector \eqref{J-phi-algebra}, can be considered as primary fields of weight one with  the most natural commutation relation,
\bea\label{WWdelta}
\{\cW^{\pm}(\phi), {\cW^{\mp}}(\phi^\prime)\}= \frac{6\pi}{c}\partial_{\phi'}
\delta(\phi-\phi')\,.
\eea
The above parallels \eqref{J-phi-algebra} for the black hole sector.\footnote{Equation \eqref{WWdelta} suggests that $\cW(\phi)$ may be viewed as the momentum conjugate to another field, e.g.~called $U(\phi)$, i.e. $U(\phi)=\int^\phi \cW$. As an intriguing and notable observation, one may readily check using \eqref{J-L} and \eqref{W-nu-def} that $L(\phi)=-\frac12\left(\frac{U'''}{U'}-\frac32\frac{U''^2}{U'^2}\right)$. That is, the energy momentum tensor of the theory in terms of $U$-fields is given by a pure Schwartzian derivative. This may be contrasted with $L$ in terms of the $h$ field, \eqref{L-h-J0} \label{W-Schawrtz} in which there is also a zero mode contribution $J_0^2$ describing `hard' degrees of freedom.}
Upon Fourier expansion,
\be\label{W-field-opt-algebra}
\cW({\varphi}) =\sum_{n\in\mathbb{Z}} \cW_n^\nu e^{i(n\pm\nu){\varphi}},\qquad \cW_n^\nu=\frac{6}{c} \langle \bcW_n^\nu\rangle\,,
\ee
the Poisson-bracket in \eqref{WWdelta} leads to the following commutator,
\begin{align}
[\bcW_n^{\pm\nu}, \bcW^{\mp\nu'}_m]
=\frac{c}{12} (n\pm\nu)\ \delta_{n,-m} 
\delta_{\nu,\nu'}\;,\qquad \nu,\nu'\in[0,1]\,,\label{Wnu-algebra}
\end{align}
which reproduces \eqref{Wnu-algebra1} when $\nu=\nu'$ (usage of $\delta_{\nu,\nu^\prime}$ is well justified in view of the discussion in the next subsection where $\nu$ will be assumed to be quantized). The result \eqref{Wnu-algebra} justifies (in hindsight) the specific factors chosen in \eqref{Wnu-algebra1}.

In other words, there is a weakly coupled field theory for $\bcW$-fields where the possible interactions with the $\bJ$-sector are subdominant in the large $c$ limit.  This has been made manifest and will be employed in the particle/black hole correspondence of section \ref{sec:two-desc}. For related discussions on quantization of conic spaces see \cite{Raeymaekers:2014kea}.

\subsection{Quantization of conic deficit angle and recovery of the near horizon algebra} 

Before moving on to constructing the Hilbert space associated with $\bcW_n^\nu$, we would like to note that from quantum gravity effects \cite{Maldacena:2000dr, David:2002wn}, it is expected that $c$ is an integer \cite{Witten:2007kt} and $\nu$ is quantized in units of $1/c$, i.e.
\be\label{nu-quantization}
\nu=\tfrac{1}{c}\,,\tfrac{2}{c}\,,\cdots\,, 1\,.
\ee
We have excluded the $\nu=0$ case which corresponds to massless BTZ. Similar quantization of the ``twist phase'' (our $\nu$ parameter) has been discussed in \cite{Banados:1999ir}.

We stress that since our discussions here is at the semi-classical level the quantization is an input, while it can be an outcome at quantum gravity level. In the known examples of consistent AdS$_3$ quantum gravity, those which have a string theory realization, e.g.~in terms of D1-D5 systems, $c$ is quantized as it is related to the product of number of D-branes and \eqref{nu-quantization} comes about due to spectral flow between the NS and R vacua \cite{Maldacena:2000dr} (see also footnote \ref{spectral-flow-footnote}). From a different viewpoint, string theory can resolve the conic singularity of the conic deficits only when the deficit angle $\nu$ is quantized as in \eqref{nu-quantization}. 
As a rough semi-classical argument for the quantization \eqref{nu-quantization} let us recall the metric for a static conic space:
\be
\extd s^2=-(1+\frac{r^2}{\ell^2})\extd t^2+\frac{\extd r^2}{1+\frac{r^2}{\ell^2}}+r^2\extd\vp^2\,,\qquad \vp\in[0,2\pi\nu]\,.
\ee
Then the smallest $\nu$ which can be semi-classically resolved is associated with $r\vp\sim \ell_{Pl}$ for $r\sim \ell$, or $\nu\gtrsim \ell_{Pl}/\ell\sim 1/c$. 
That is, our Bohr-type quantization condition  \eqref{nu-quantization} is clever enough to know the essential information about the spectrum of the primaries of the theory.

To facilitate building the Hilbert space, and especially noting the hermiticity condition \eqref{W-dagger-W-conjugates}, we can collectively gather all $\bcW_n^\nu$ into Fourier modes $\bcJ_n$ of a single field $\bcJ$, 
\be\label{J-j-map}
\bcJ_{c(n\pm\nu)} \equiv {\sqrt{6}}\bcW_n^{\pm\nu} \,.
\ee
The commutation relation \eqref{Wnu-algebra}  then takes the simple form of a $\hat u(1)$ current algebra
\be\label{bcJ-algebra}
[\bcJ_n,\bcJ_m]=\frac{n}{2}\delta_{n,-m}\,,
\ee
and the hermiticity condition \eqref{W-dagger-W-conjugates} implies 
\be
\bcJ_n^\dagger=\bcJ_{-n}\,. 
\ee
See appendix \ref{appendix-D} for more discussions on hermitian conjugation and unitarity. The identification \eqref{J-j-map} implies that the center element of the algebra is $\bcJ_0=\sqrt{6}\bcW_{1}^{\nu=-1}=\sqrt{6}\bcW^{\nu=1}_{-1}$. Moreover, all $\bcJ_{nc}$ generators would correspond to $\nu=1$, and are associated with global AdS$_3$ while the $\bcJ_{c(n+\nu)}$ generators with $\nu\neq 1$ (or $\bcW^{\nu\neq1}_n$) represent conic spaces.

There is a redundancy in the identification \eqref{J-j-map} under the simultaneous shifts: $\pm\nu\rightarrow\pm\nu\pm1$ and $n\rightarrow n\mp1$. We use this redundancy to relate $\bcW$'s with positive and negative $\nu$; explicitly,
\be\label{W-J-negative-nu}
\bcW_n^{\pm(1-\nu)}= \bcW^{\mp\nu}_{n\pm 1}\,.
\ee
We further deduce
\begin{align}
[\bcW_n^{\pm\nu}, \bcW^{\pm\nu'}_m]=\frac{c}{12} (n\pm\nu)\delta_{n+m,\mp1}\delta_{\nu+\nu',1}
\qquad \nu,\nu'\in(0,1]\,.\label{Wnu-algebra2}
\end{align}
Unlike the commutator \eqref{Wnu-algebra}, which is also true for $\nu=0$ here we have $\nu\in(0,1]$. This means that all the generators with negative $\nu$ can be mapped to those with positive $\nu$ and hence each operator in the set of $\bcW_n^{\pm\nu}$ with $n\in \mathbb{Z}$, $\nu\in(0,1]$ appears twice. To avoid this doubling we choose to work with $\bcW_n^\nu,\ \bcW_{-n}^{-\nu},\ n\geq 0, \nu\in(0,1]$ as independent generators.

The algebra \eqref{bcJ-algebra} is basically the same as the algebra of $\bJ_n$ \eqref{J-algebra} since it is related to it upon a redefinition and rescaling of the generators. 
However, physically these algebras are distinct. In particular, the algebra \eqref{bcJ-algebra} does not involve any information about AdS$_3$ (like the AdS-radius $\ell$ contained in the central charge $c$) or the black hole background. In fact, it has been argued \cite{Afshar:2016kjj} that one may obtain \eqref{bcJ-algebra} as the asymptotic/symplectic algebra of the Rindler space \eqref{eq:Rindler}. This latter provides a strong support for the quantization scheme adopted in \eqref{Wnu-algebra}.

\subsection{Hilbert space of near horizon soft hairs}

The unitary representations of the algebra \eqref{bcJ-algebra} can be constructed in the same manner as that of \eqref{J-algebra}. We define a unique vacuum state through
\be\label{bcJ-vacuum}
\bcJ_n|0\rangle=0,\ \  \forall n\geq 0.
\ee
Note that in comparison to the black hole case of $\cH_{\textrm{\tiny BTZ}}$ we have chosen the vacuum state (corresponding to the global AdS$_3$) to have vanishing $\cJ_0$.\footnote{Recall also the discussions below \eqref{Wn-states-J0} and the fact that $|0\rangle$ and vacuum states of the $\bJ$ algebra $|0;J_0\rangle$ should be different and in principle belong to two different Hilbert spaces.} Recalling that the algebra of $\bcJ_n$ and $\bJ_n$ are the same, the choice of $\bcJ_0=0$ for vacuum may be attributed to working in the near horizon soft hair sector \cite{Afshar:2016uax, Sheikh-Jabbari:2016npa}. 
Before continuing we make a notational alert. Since by assumption $\nu$ is quantized in units of $1/c$ hereafter we use the notation
\be\label{Wr-n}
\bcW^r_n\equiv \bcW^\nu_n \qquad \text{for} \qquad \nu c\equiv r=1,\cdots,c\,.
\ee
In terms of the $\bcW^r_n$ generators, the vacuum conditions \eqref{bcJ-vacuum} may be written as
\be\label{bcW-vacuum}
\bcW_n^r|0\rangle=0,\ \  \forall n \geq 0\,,\;  r\geq 1\,.
\ee
The condition $\cJ_0=0$ means that we are working in the soft hair sector. (Unlike the black hole case a vacuum state with non-zero $\cJ_0$ would not make sense.)
The Hilbert space of near horizon soft hairs can then be constructed as
\be\label{NH-soft-hair}
|\Psi(\{n_i\})\rangle=\prod_{n_i>0} \bcJ_{-n_i} |0\rangle.
\ee
Recalling the identification \eqref{J-j-map}, one can then readily show that the Hilbert space of near horizon soft hairs is exactly the same as $\cH_{\textrm{\tiny CG}}$ for quantized values of $\nu$ \eqref{nu-quantization}. Hereafter, we will only focus on such quantized $\nu$'s, and by $\HCG$ we mean states corresponding to such conic spaces. As discussed, recalling the definitions of $\bcW$ fields and $\bcJ_n$, \eqref{bcW} and \eqref{J-j-map}, states of the form \eqref{NH-soft-hair} describe generic coherent states of particles on AdS$_3$ (i.e. conic spaces and their Virasoro descendants).   Note that 
only single particle states in the above, the $\bcJ_{-n}|0\rangle$ states, have a straightforward geometric interpretation.
In particular, if both left and right sectors are associated with $\bcW^{-r}_0|0\rangle$ states, they correspond to conic space or global AdS$_3$ geometries \cite{Sheikh-Jabbari:2016npa,Sheikh-Jabbari:2016unm}. 

\subsection{Virasoro algebra of $\bcW$-fields} We can construct Virasoro generators in terms of the free fields of twisted boundary conditions $\bcW_n^r$ as follows. Let us define 
\begin{align}\label{Ln-r} 
\bL_n^r&=\frac{6}{c}\sum_{p\in \mathbb{Z}} :\!\! \bcW^{-r}_{n-p}\bcW_p^r\!:+ \frac12 f_r\delta_{n,0}\nn\\
&= \frac{6}{c} \left( \sum_{p\leq -1} \bcW_p^r \bcW_{n-p}^{-r} +\sum_{p>-1} \bcW_{n-p}^{-r} \bcW_p^r \right) +\frac12 f_r \delta_{n,0}
\end{align}
with $r=1,2,\cdots, c$ and \cite{Polchinski:1998rq}
\be\label{zero-point-en}
f_r\equiv\sum_{n=1}^\infty\left(n-\tfrac{r}{c}\right)\simeq\frac{1}{24}-\frac{1}{8}\left(\tfrac{2r}{c}-1\right)^2,
\ee
where we used zeta-function regularization to evaluate the infinite sum to extract its unique finite part.
We note that  the $r$ superscript on $\bL_n^r$ does not correspond to twisted boundary conditions on $\bL^r(\phi)\propto\sum_n \bL_n^r e^{in\phi}$; instead $\bL^r(\phi+2\pi)=\bL^r(\phi)$ for any $r$. One may then show that
\be\label{Ldagger-L1-r}
\bL_n^{-r}=\bL_n^r+\frac{r}{2c}\delta_{n,0},\qquad (\bL_n^{r})^\dagger=\bL_{-n}^{r}\,,\qquad \bL_n^{c-r}=\bL_n^{r},
\ee
and that 
\begin{align}\label{Lrn-W-commutation}
[\bL_n^{s}, \bcW^{\pm r}_{m}] &= -\frac12\left(m\pm \tfrac{r}{c}\right)\bcW^{\pm r}_{m+n}(\delta_{r,s}+\delta_{r,c-s})\,,\\
[\bL_n^{r}, \bL_m^{s}] &=\frac12\left[(n-m) \bL_{n+m}^{r}+\left(mf_r+\tfrac{1}{2}\sum_{p=0}^{-1+m}\left(p-m+\tfrac{r}{c}\right)\left(p+\tfrac{r}{c}\right)\right)\delta_{m+n,0}\right](\delta_{r,\,s}+\delta_{r,c-s})\nn\\
&=\frac12\left[(n-m) \bL_{n+m}^{r}+\frac{1}{12}n^3\delta_{m+n,0}\right](\delta_{r,\,s}+\delta_{r,c-s}),
\label{Lrn-Vir-algebra1}
\end{align}
where we fixed the zero-point energy $f_r$ as in \eqref{zero-point-en} such that \eqref{Lrn-Vir-algebra1} is a Virasoro algebra of central charge one (associated with twisted modes) preserving the vacuum. This value coincides with \eqref{zero-point-en}.  In computing the above commutators we have extensively used \eqref{Wnu-algebra}, \eqref{W-J-negative-nu} and \eqref{Wnu-algebra2} or in more explicit form:
\begin{align}
[\bcW_n^r, \bcW^s_m] &=\frac{c}{12}\left(n+\tfrac{r}{c}\right)\delta_{m+n,-1}\,\delta_{r+s,c}\,,\qquad
[\bcW_n^r, \bcW^{-s}_m]=\frac{c}{12}\left(n+\tfrac{r}{c}\right)\delta_{m+n,0}\,\delta_{r,s}\,\\
[\bcW_n^{-r}, \bcW^{-s}_m] &=\frac{c}{12}\left(n-\tfrac{r}{c}\right)\delta_{m+n,1}\,\delta_{r+s,c}\,,\qquad 
\bcW_n^{r}=\bcW_{n+1}^{r-c}\,,\qquad\qquad\bcW_n^{-r}=\bcW_{n-1}^{-r+c}\,.
\end{align}

The commutator \eqref{Lrn-W-commutation}, noting that the coefficient on the right hand side is $n$ independent, implies that $\bcW_n^r$ is a primary operator of conformal weight one, while the $r/c$ part in $m+r/c$ reflects the twisted boundary condition of the $\cW$ field  \eqref{W-periodicity}.
Additionally, observe that  $\bcW_{-1}^{r}|0\rangle=\bcW^{-(1-r)}_{0}|0\rangle$ are primary states of weight $1-r/c$ and hence the unitarity bound implies $r\leq c$. This is of course compatible  with our earlier discussions that $\nu=r/c\leq 1$. We also note in passing that, using the commutators \eqref{J-W-commutator}, one can show that $[\bJ_0,\bL_n^{r}]=0$. This will be important for our construction of the black hole microstates. 

Using the $\bL_n^{r}$ we can define a Virasoro algebra at central charge $c$ by summing over all allowed values of the integer $r$:
\be\label{Ln-sum-Ln-r}
\bL_n=\sum_{r=1}^{c}\bL_n^{r}=\frac{1}{c}\sum_{p\in\mathbb{Z}} :\!\! \bcJ_{nc-p}\bcJ_p\!\!:-\frac{1}{24c}\delta_{n,0}
\ee
yielding
\be\label{Vir-algebra-HCG-sector}
[\bL_n, \bL_m]=(n-m) \bL_{m+n}+ \frac{c}{12}n^3\delta_{m,-n}\,, \qquad[\bL_n,\bcW_m^r]=-\left(m+\tfrac{r}{c}\right)\bcW_{n+m}^r\,.
\ee
Thus, the  $\bL_n$ generate a Virasoro algebra at central charge $c$ and the $\bcW_m^r$ are primaries with respect to these Virasoro generators. To arrive at \eqref{Vir-algebra-HCG-sector} we have used \eqref{J-j-map} and the fact that
\be
\sum_{r=1}^c f_r=-\frac{1}{12c}\,.
\ee

Given the algebra \eqref{Vir-algebra-HCG-sector} one can readily check that the vacuum state $|0\rangle$, for which $\bL_n^r|0\rangle =0,\ n>0$, is not an $SL(2,\mathbb{R})$ invariant vacuum. However, starting from $\bcJ_n$ and using the usual Sugawara construction yields a Virasoro algebra at central charge one with generators $\bcL_n$. This latter has been carried out in appendix \ref{app:Banados-map} and as shown there, $\bL_n=\frac{1}{c}(\bcL_{nc}-\frac{1}{24}\delta_{n,0})$.
The vacuum state $|0\rangle$ is the $SL(2,\mathbb{R})$ invariant vacuum state in this algebra, as $\bcL_n|0\rangle=0,\ n\geq -1$. 

To summarize, we have given two descriptions for the Virasoro algebra
at the Brown--Henneaux central charge, \eqref{twisted-sugawara} and \eqref{Ln-sum-Ln-r}. In the next section we  exploit the relation between these two descriptions and propose a black hole/particle correspondence that lies at the heart of the horizon fluff proposal.

\section{Black hole/particle correspondence and the horizon fluff proposal}\label{sec:two-desc}

In the previous section we noted that in the canonical description we have the family of black holes and the family of conic plus global AdS, respectively associated with $\HBH$ and $\HCG$. Similar two classes may of course be recognized in the microcanonical description. Moreover, we argued that there are two different descriptions, one in terms of $\bJ$ fields and the other in terms of their coherent states $\bcJ$'s. The key assumption promoted in this section is that both descriptions are equivalent. 

In section \ref{sec:BH-CG-duality} we state the black hole/particle correspondence that is suggested by our analysis in the previous sections and relate the operators in the black hole sector, $\bJ$, with the operators in the particle sector, $\bcJ$. In section \ref{sec-5} we use this correspondence to revisit the horizon fluff proposal \cite{Afshar:2016uax, Sheikh-Jabbari:2016npa}. The main improvement over the original proposal is that we do not have to assume a Ba\~nados-type map (see appendix \ref{app:Banados-map}) between Virasoro generators; rather, we can derive this relation  from our set of assumptions that we have spelled out above. The actual counting of our BTZ microstates, recalled in section \ref{se:hardyology}, is then straightforward and works exactly as in previous work.

\subsection{Two free fields and black hole/particle correspondence}\label{sec:BH-CG-duality}

As discussed in section \ref{canonical-subsection}, the canonical description is naturally described in terms of the field $J(\phi)$. The discussions in section \ref{sec:can-sympl} and especially the commutation relation \eqref{J-phi-algebra} indicate that $J(\phi)$ has a natural interpretation in terms of  the momentum conjugate to a free field, which was denoted by  $\Phi$ in \eqref{Phi-field}. Since, 
\be
[\Phi(\phi_1), \Pi_\Phi(\phi_2)]= 2\pi i  \delta(\phi_1-\phi_2), \qquad \Pi_\Phi(\phi)=-\frac{c}{12} J(\phi)
\ee
the pair $(\Phi, \Pi_\Phi)$ provides a good canonical pair for real $J_0$ [recalling the periodicity property \eqref{Phi-periodicity}], i.e., the case of black holes. For this case Fourier modes of the energy-momentum tensor of the corresponding field theory are given in terms of $\bL_n$ \eqref{twisted-sugawara}. In this sector the operators $\bJ_n, \bL_n$ are defined on $\HBH$.

For the case of conic spaces it is desirable to provide another natural free field. As discussed in section \ref{sec:more-on-HCG}, the primary field of twisted periodicity $\cW(\phi)$ is the natural choice. The Fourier modes of the energy momentum tensor for this sector are given in \eqref{Ln-sum-Ln-r}. In this sector, as discussed, the operators $\bcW_n^r, \bL_n$ are defined on $\HCG$. As mentioned the $\bcW$ operators may be viewed as non-local functions of $\bJ$, cf.~\eqref{W-nu-def}. In other words, states in $\HCG$ are composite states in $\HBH$ and vice-versa.
This means that black holes may be viewed as composite states in $\HCG$. We will use and exploit this to identify the black hole microstates. 

Having two different descriptions of the same energy momentum tensor together with the discussions above suggests that the two expressions for $\bL_n$, \eqref{twisted-sugawara} and \eqref{Ln-sum-Ln-r}, provide two dual descriptions for the same theory. This correspondence is similar in spirit to  Sine-Gordon/Thirring model duality where solitons of the former appear as fundamental degrees of freedom of the latter and vice-versa \cite{Coleman:1974bu, Mandelstam:1975hb}. The statement of our correspondence \emph{in the leading order in central charge $c$} is then obtained from equating \eqref{twisted-sugawara} and \eqref{Ln-sum-Ln-r}. That is,
\be\label{duality-statement}
\frac{1}{c}\sum_{p\in\mathbb{Z}} \colon\!\bcJ_{nc-p}\bcJ_p\!\colon=in \bJ_n+ \frac{6}{c}\sum_{p\in\mathbb{Z}}  \colon\!\bJ_{n-p}\bJ_p\!\colon
\ee

As the zeroth order evidence we note that both sides in the correspondence \eqref{duality-statement} generate a Virasoro algebra at central charge $c$. Next, we note that the commutator of both sides with $\bcW^r_n$ fields leads to the same result. The commutator of the left hand side is given in \eqref{Vir-algebra-HCG-sector}. The one on the right hand side is the quantized version of  \eqref{eq:primary}.
As further evidence we note that both sides of \eqref{duality-statement} commute with $\bJ_0$. This latter will be crucially used in our discussions of the microstates in the next subsection. In particular, the physical charges of black hole like mass and angular momentum are specified by $\bJ_0$ eigenvalues and --- as discussed and stressed in \cite{Sheikh-Jabbari:2016npa,Compere:2015knw, Sheikh-Jabbari:2016unm} --- entropy should be an invariant of Virasoro coadjoint orbits, and hence can only be a function of $J_0^\pm$. 

Finally, we justify the name \emph{near horizon soft hair} for the states of the form \eqref{NH-soft-hair}. Namely, note that the algebra of $\bJ_n$ and $\bcJ_n$ are essentially the same, and that $J_0$ is the energy as measured by the near horizon observer. Thus, what remains to be shown as a justification is to show that the vacuum state $|0\rangle$ is the same as the vacuum state $|0;J_0=0\rangle$. This can be readily verified by acting with both sides of \eqref{duality-statement} on $|0\rangle$ and noting that $|0;J_0=0\rangle$ is the only state on which the right hand side of \eqref{duality-statement} vanishes.

\begin{figure}
    \centering
    \includegraphics[width=15cm]{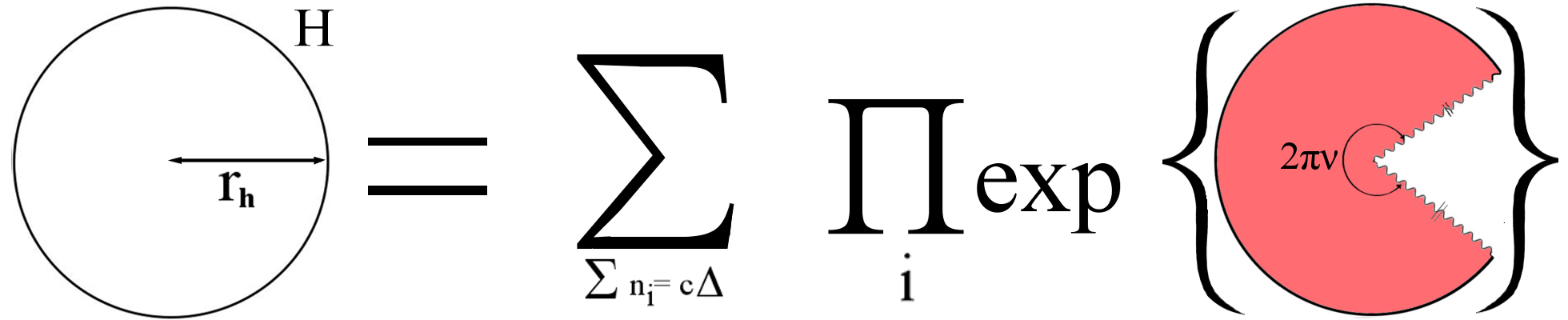}
    \caption{Schematic depiction of horizon fluff proposal}
    \label{horizon-fluff-fig}\label{fig:4}
\end{figure}
Schematically our horizon fluff proposal is displayed in figure \ref{horizon-fluff-fig}. A BTZ black hole (left figure) is a collection of coherent states of particles on AdS$_3$ (the conic spaces and their Virasoro descendants). The exponentiation in the right figure is to demonstrate dealing with coherent states. This coherent states are among soft hairs, as they have vanishing $\bcJ_0$ and their `near horizon energy' is equal to $c\Delta$, related to mass and angular momentum of the BTZ black hole through \eqref{eq:vevs} below.

\subsection{Horizon fluff proposal, revisited and tested}\label{sec-5}

The correspondence equation \eqref{duality-statement} relates the $\HBH$ and $\HCG$ sectors of the Hilbert space ${\cal H}_{\textrm{\tiny Vir}}$. In particular, one can view it as an equation which specifies a given AdS$_3$ black hole state (which is in general a coherent state in $\HBH$ \cite{Sheikh-Jabbari:2016unm}) in $\HCG$. For a given black hole state there is a sector of $\HCG$ which solves \eqref{duality-statement}. This sector is exactly what we identify as black hole microstates. 

Our proposal is now realized in a very simple way. Let us denote the BTZ black hole of mass $M$ and angular momentum $J$ by $|0;J_0^+\rangle \otimes |0;J_0^-\rangle$ where
mass $M$ and angular momentum $J$ satisfy the relations
\be
\langle \bL_n^\pm\rangle\big|_{\hspace*{-2mm}\vspace*{-6mm}\begin{array}{c}\textrm{\tiny BTZ}\vspace*{-3mm}\\ \textrm{\tiny can.}\end{array}}={\frac{c}{6}}(J_0^\pm)^2\delta_{n,0}\,,\qquad \Delta_\pm=\frac12(\ell M\pm J)={\frac{c}{6}} (J_0^\pm)^2\,.
\label{eq:vevs}
\ee
The subscript ``BTZ can.'' refers to the fact that the above provides a definition of the BTZ black hole in our canonical description, cf.~section \ref{canonical-subsection}. In this canonical description we are dealing with diffeomorphisms and associated charges on a background black hole of fixed horizon temperature, while the expectation value of charges and in particular $\bJ_0^\pm$ specify the mass and angular momentum of the black hole. We shall return to this fact in the next section when we compute the log-corrections.

Microstates of this black hole, the ``horizon fluff'' \cite{Afshar:2016uax, Sheikh-Jabbari:2016npa}, are the set of all states in $\cH_{\textrm{\tiny CG}}$ that satisfy the duality relation \eqref{duality-statement}. If we denote them by $|\cB(\{n_i^\pm\});J^\pm_0\rangle$, they satisfy \cite{Afshar:2016uax, Sheikh-Jabbari:2016npa}
\be\label{BTZ-microstate-equation}
\langle \cB'(\{n_i^\pm\});J^\pm_0|\  \bL_n^\pm\ |\cB(\{n_i^\pm\});J^\pm_0\rangle={\frac{c}{6}}(J_0^\pm)^2\delta_{\cB,\cB'}\delta_{n,0}\,,
\ee
and hence
\be\label{BTZ-microstates}
|\cB(\{n_i^\pm\}); J^\pm_0\rangle = \prod_{\{n_i^\pm>0\}} \!\!\!\!\big(\bcJ_{-n_i^+}^+ \cdot \bcJ_{-n_i^-}^-\big) |0\rangle \,,\quad \text{such that}\quad \sum n_i^\pm= c\Delta^\pm\,,
\ee
or linear combinations thereof. 

Note that $|\cB(\{n_i^\pm\}); J^\pm_0\rangle$ involves $n_i<c$ as well as $n_i>c$ states. The states may be viewed as a collection (or condensate) of coherent states of particles on AdS$_3$ or their Virasoro excitations. The set of all states $|\cB(\{n_i^\pm\}); J^\pm_0\rangle$ form a vector space over $\cH_{\textrm{\tiny CG}}$. A schematic presentation of the horizon fluff has been shown in figure \ref{horizon-fluff-fig}.

We note that the discussion above, valid for BTZ black holes, can be extended to a generic AdS$_3$ black hole, which is a Virasoro descendant of generic BTZ black holes, with the result that the number of microstates (just like the horizon temperature and angular velocity) is independent of the Virasoro excitation, see \cite{Sheikh-Jabbari:2016npa}. This is of course expected since the laws of black hole thermodynamics are orbit-invariant relations \cite{Sheikh-Jabbari:2016unm}. In other words, in the figure \ref{horizon-fluff-fig}, one can replace the figure on the left with the picture with wavy line horizon in figure \ref{fig:1}.

\subsection{Microstate counting and black hole entropy} \label{se:hardyology}

Given the BTZ black hole microstates \eqref{BTZ-microstates}, we can count them and obtain the entropy. The number of microstates for a BTZ black hole of given $\Delta_\pm$ is  the number of partitions $p(c\Delta_\pm)$ into non-negative integers. For large $c\Delta_\pm$ one may use the Hardy--Ramanujan formula to evaluate $p(c\Delta^\pm)$ (see e.g.~\cite{Carlip:1998qw} and Refs.~therein). The details of the computation may be found in \cite{Afshar:2016uax, Sheikh-Jabbari:2016npa}. The entropy is the logarithm of the number of microstates,
\be\label{microstate-counting}
 S^{\textrm{\tiny fluff}} = \ln p(c\Delta^+) + \ln p(c\Delta^-)= 2\pi \Big(\sqrt{\frac{c\Delta^+}{6}} +  \sqrt{\frac{c \Delta^-}{6}}\Big) -\ln (c\Delta^+)-\ln(c\Delta^-)+\cdots 
\ee
where we have assumed $c\Delta^\pm\gg 1$, and $\cdots$ denotes sub-leading terms in $c\Delta^\pm$. The first, leading terms in \eqref{microstate-counting} is nothing but the standard Cardy formula which reproduces the BTZ black hole Bekenstein-Hawking entropy \cite{Strominger}. The above, however, has logarithmic corrections that we analyze in the next section.

\section{Horizon fluff and logarithmic correction to  black hole entropy}\label{sec:log-correction}

In previous sections we have presented two descriptions of the geometries in \eqref{afshar2}, a canonical description discussed in sections \ref{canonical-subsection} and \ref{sec-canonical-Hilbert}, and  a microcanonical description in section \ref{microcanonical-subsection}. We have shown above that to leading order our horizon fluff proposal reproduces the  Bekenstein--Hawking area law \eqref{eq:BH}, which is a semi-classical consistency check. As pointed out in the introduction (see there for references) also the subleading logarithmic corrections are accessible by semi-classical means and therefore provide a non-trivial test of our proposal. In particular, our horizon fluff proposal should be able to reproduce the correct numerical coefficient $N_{\textrm{\tiny log}}$ in \eqref{eq:logS}. The purpose of this section is to check whether or not this is the case.

There are generally two types of contributions to $N_{\textrm{\tiny log}}$. One (type 1) is a genuine 1-loop correction where e.g.~zero modes contribute. The other (type 2) is a contribution that depends just on the ensemble that one uses to express the entropy. In the pertinent case of BTZ black holes with Brown--Henneaux boundary conditions the former contribution vanishes, while the latter vanishes in the canonical ensemble.\footnote{By ``canonical'' we mean the ensemble where temperature and angular potential are fixed; by microcanonical we mean that the energy and angular momentum are fixed. In this section we focus only on one chiral sector, so that we have only one charge and one associated chemical potential. The log-corrections, similarly to the leading order part, are expected to be the sum of left and right movers (e.g. see \cite{Loran:2010bd}).} Transforming from canonical to microcanonical ensemble leads to the famous contribution of $\tfrac 32$ to $N_{\textrm{\tiny log}}$ \cite{Carlip:2000nv, Loran:2010bd}:
\eq{
S_{\textrm{\tiny can}}^{\textrm{\tiny BTZ}} = S_{\textrm{\tiny BH}} + \dots, \qquad S_{\textrm{\tiny mic}}^{\textrm{\tiny BTZ}} = S_{\textrm{\tiny can}}^{\textrm{\tiny BTZ}} - \tfrac32\,\ln S + \dots 
}{eq:log1}
Note that under the log we refrain from labelling the entropy, since to the order of interest it does not matter which entropy is used here, $S_{\textrm{\tiny BH}}$, $S_{\textrm{\tiny can}}$, $S_{\textrm{\tiny mic}}$ or some other entropy that coincides with the Bekenstein--Hawking result up to subleading corrections. The result \eqref{eq:log1} is also obtained microscopically from a Cardy-type of computation \cite{Carlip:2000nv, Loran:2010bd} where the  modular invariance yields a high/low temperature duality for the partition function.

The analysis of section \ref{sec-5} established the result \eqref{microstate-counting} which may also be written as
\eq{
S^{\textrm{\tiny fluff}} = S_{\textrm{\tiny BH}} - 2\,\ln S + \dots ,
}{eq:log10}
differing from the log contributions contained in \eqref{eq:log1}. To resolve the apparent discrepancy, we note the key issue that our microstate counting was based on the charges defined in what we called a canonical description of the black hole. That is, as discussed in section \ref{canonical-subsection}, we defined the charges $\bJ_n$ by choosing the boundary conditions of metric fluctuations to those keeping the horizon temperatures fixed. However, in sections \ref{sec-canonical-Hilbert} and \ref{sec:two-desc} we defined the state of BTZ black hole with the expectation value of $\bJ_n$ over a state for which the mass and angular momentum is fixed, which corresponds to the microcanonical ensemble. Explicitly, in our canonical description $\langle \bJ_n\rangle_{\textrm{\tiny can}}= cJ_0/6\; \delta_{n,0}$ for the BTZ black hole. [See also the discussion below \eqref{eq:vevs}.] Therefore, our description in terms of $\bJ_n$ is a hybrid of microcanonical and canonical descriptions. 

In the next two subsections we disentangle our hybrid construction and then re-derive the result \eqref{eq:log10} in two conceptually quite different ways. 
In section \ref{se:log1}, we resolve the issue by modifying the hybrid description in favor of a fully microcanonical one. That is, using \eqref{J-h-Phi} we map the canonical part of the hybrid description, the $\bJ_n$, to their microcanonical counterpart. This is mostly a type-1 derivation.
In section \ref{se:log2}, we take a different route to verify the log-corrections of the fluff proposal. This subsection is based on the S-type of dualities for theories with Lifshitz-type of scaling for $z=0$ and $z=1$ discussed in \cite{Perez:2016vqo, Grumiller:2017jft} and on  our black hole/particle correspondence (cf.~section \ref{sec:BH-CG-duality}) to derive \eqref{eq:log10}, so it is exclusively a type-2 derivation.
The final results in both sections agree with each other and with the results \eqref{eq:log10} and \eqref{eq:log1}.

\subsection{From hybrid to purely microcanonical ensemble,
log-corrections from replica trick}\label{se:log1}

In this subsection by the subscript BTZ we exclusively mean the microcanonical ensemble associated with a BTZ black hole (in contrast with canonical or hybrid description).  The microcanonical entropy is obtained through replacing $\texttt{J}_0$ in \eqref{eq:intro1}  with 
\eq{
\texttt{J}_0 \to \Big\langle \frac{1}{2\pi} \int\limits_0^{2\pi}\extd\phi\, \bJ(\phi)\Big\rangle_{\textrm{\tiny BTZ}}
}{eq:log11}
where $\langle\bJ(\phi)\rangle=\frac{c}{6}J(\phi)$ (\emph{cf.} footnote \ref{operator-J}). With the above the microcanonical fluff entropy \eqref{microstate-counting} or \eqref{eq:log10} reads as
\be\label{S-fluff-mic}
S^{\textrm{\tiny fluff}}_{\textrm{\tiny mic}}=  \Big\langle\int\limits_0^{2\pi}\extd\phi\,  \bJ(\phi)\Big\rangle_{\textrm{\tiny BTZ}}-2 \ln \left(\Big\langle\int\limits_0^{2\pi}\extd\phi\,  \bJ(\phi)\Big\rangle_{\textrm{\tiny BTZ}}\right)+\cdots 
\ee
To compute the above we use the analysis in section \ref{sec:can-to-mcan} and \eqref{J-h-Phi} to map the $J$ fields onto the microcanonical fields $h$, which in the quantum version reads as
\eq{
\bJ(\phi)=\frac{c}{6}J_0 \bh'(\phi)-\frac{1}{2} \frac{\bh''(\phi)}{\bh'(\phi)}\,.
}{eq:log20}
In writing the second term in principle one needs to assume a specific ordering. However, the leading order in the final result which we are interested in here does not depend on this ordering. One may integrate both sides and take the expectation value over the BTZ state:
\eq{
\Big\langle \int\limits_0^{2\pi}\extd\phi\, \bJ(\phi)\Big\rangle_{\textrm{\tiny BTZ}}=\frac{c}{6}J_0 \big\langle \bh(\phi)\big|_0^{2\pi}\big\rangle_{\textrm{\tiny BTZ}}-\frac{1}{2}\, \big\langle  \ln(\bh'(\phi))\big|_0^{2\pi}\big\rangle_{\textrm{\tiny BTZ}}. }{eq:log19}
Recalling the definition of $\bh(\phi)$ and discussions of sections \ref{sec-canonical-Hilbert} and \ref{sec:two-desc}, we learn that $\langle \bh(\phi)\big|_0^{2\pi}\rangle_{\textrm{\tiny BTZ}}={2\pi}$.

We need to compute the log-term. We first note that the log-term in \eqref{eq:log19} is classically zero due to the periodicity of $h'(\phi)$ and hence a nonvanishing contribution from this log-term is a quantum effect, it is a 1-loop result (hence type-1).
To this end, since we are only interested in the leading log-correction we can safely work in the large-$J_0$ limit, where $h(\phi)$ provides a convenient free-field representation with the Hamiltonian \eqref{L-h-J0} $L(\phi)\simeq J_0^2 h'(\phi)^2$. We use the replica trick to convert the log-term into a monomial,
\eq{
 \big\langle  \ln(\bh'(\phi))\big|_0^{2\pi}\big\rangle_{\textrm{\tiny BTZ}}=\frac{\extd}{\extd n}  \big\langle  (\bh'(\phi)^n)\big|_0^{2\pi}\big\rangle_{\textrm{\tiny BTZ}}\big|_{n=0}.
}{eq:angelinajolie}
The above, in the large-$J_0$ limit is a straightforward free field theory computation of an $n$-point function in the pinching limit. One can regularize it using the point-splitting method. Explicitly, \eqref{eq:log20} may be conveniently written as
\eq{
\boldsymbol{h}' \simeq {\frac{6}{cJ_0}}\,\big(\boldsymbol{\Sigma}(\phi) + \boldsymbol{J}_0\big),
}{eq:hprime}
where the quantity $\boldsymbol{\Sigma}(\phi)$ is defined by
\eq{
\boldsymbol{\Sigma}(\phi) =  \sum_{p>0} \,\big(\boldsymbol{J}_p e^{ip\phi} + \boldsymbol{J}_p^\dagger e^{-ip\phi}\big),
}{eq:sigma}
The regularized $n$-point function of $\boldsymbol{h}^\prime$ may be written as 
\eq{
\big\langle(\boldsymbol{h}^\prime)^n\big|_0^{2\pi}\big\rangle_{\textrm{\tiny BTZ}} = {\rm const.} + \frac{1}{J_0^n}\,F_n(\phi_1,\dots,\phi_n) + \dots
}{eq:npt}
where $\phi_i-\phi_{i+1}$ is a small angle (related to the point-splitting regulator) and $\phi_n-\phi_1=2\pi$. The const.~piece is an irrelevant $n$-independent constant. We introduced the definition
\eq{
F_n(\phi_1,\dots,\phi_n) = {\left(\frac{6}{c}\right)^n}\ \big\langle\boldsymbol{\Sigma}(\phi_1)\dots\boldsymbol{\Sigma}(\phi_n)\big\rangle\,.
}{eq:Fn}
The ellipsis in the $n$-point function \eqref{eq:npt} denotes terms of higher order in $n$ that do not contribute to the $n\to 0$ limit in the replica trick \eqref{eq:angelinajolie}. 
To   compute $F_n$ for generic large $n$ we use the above mentioned point-splitting and then analytically continue the result to $n\to 0$, yielding $F_0=1$.
Insertion of \eqref{eq:npt} into \eqref{eq:angelinajolie} then establishes the result
\eq{
\big\langle  \ln(\bh'(\phi))\big|_0^{2\pi}\big\rangle_{\textrm{\tiny BTZ}}=-\ln J_0 +\  J_0~\text{-independent terms}.
}{eq:log18}

Plugging \eqref{eq:log19} and \eqref{eq:log18} into \eqref{S-fluff-mic} we obtain 
\be
{S_{\textrm{\tiny mic}}^{\textrm{\tiny fluff}}= }
S^{\textrm{\tiny fluff}}+\tfrac12 \,\ln S + \dots 
=S_{\textrm{\tiny BH}} - \tfrac32\,\ln S + \dots=S_{\textrm{\tiny mic}}^{\textrm{\tiny BTZ}}.
\ee
We have hence established that entropy coming from our horizon fluff proposal, once computed in a fully microcanonical description matches with the BTZ black hole microcanonical entropy including the log-corrections.

\subsection{Log corrections from  Lifshitz scaling} 
\label{se:log2} 

We present now an alternative derivation that requires only type-2 log corrections, albeit of two different sources. To this end we quote a result \cite{Gonzalez:2011nz, Shaghoulian:2015dwa, Perez:2016vqo, Grumiller:2017jft} that applies to black holes with anisotropies of Lifshitz-type with dynamical scaling exponent $z$ (so that $t\to\lambda^zt$ and $\phi\to\lambda\phi$) \cite{Shaghoulian:2015dwa}:
\eq{
S_{\textrm{\tiny mic}}^{z} = S_{\textrm{\tiny can}}^{z} - \big(z + \tfrac12\big)\,\ln S + \dots
}{eq:log13}
We need this result only for two special cases, the isotropic one $z=1$ and the Rindler one $z=0$.

The $z=1$ case corresponds to the usual BTZ black hole in Brown-Hennaux description which is dual to an isotropic CFT$_2$  and hence,  
\eq{
S^{z=1}_{\textrm{\tiny can}} = S^{\textrm{\tiny CFT}}_{\textrm{\tiny can}} = S^{\textrm{\tiny BTZ}}_{\textrm{\tiny can}} = S_{\textrm{\tiny BH}},
}{eq:log14}
whereas the $z=0$ case corresponds to the Rindler geometry  which describes the near horizon region \cite{Perez:2016vqo, Afshar:2016kjj, Grumiller:2017jft}. As argued in \cite{Grumiller:2017jft} and as evidenced by the microscopic matching, the correct scaling properties and the correct ground state energies, the leading order contribution $S^{z=0}$ indeed reproduces the desired leading order result \eqref{eq:intro1}.  Our analysis in this paper and in particular in section \ref{sec:two-desc} suggests that  this matching should extend to the subleading log corrections, 
\eq{
S^{\textrm{\tiny fluff}} = S^{z=0}_{\textrm{\tiny mic}} + \dots 
}{eq:log15}
where as before $\cdots$ stands for corrections subleading to log-corrections. Equation \eqref{eq:log15} may be viewed as an alternative reading or statement of our proposal, recalling our discussions in previous sections. Therefore, verifying that \eqref{eq:log15} indeed holds, provides another check for our proposal. That is what below we argue for. 

To this end we recall the duality statement in our horizon fluff proposal \eqref{duality-statement} and that it implies the canonical entropy in our near horizon picture should be identified with the microcanonical BTZ entropy in the usual asymptotic picture:
\eq{
S_{\textrm{\tiny can}}^{{z=0}}  = S_{\textrm{\tiny mic}}^{{z=1}}. 
}{eq:log0}
This is so, since we fixed the asymptotic eigenvalues \eqref{eq:vevs} and thus are using the microcanonical ensemble on the right hand side of \eqref{duality-statement}, but we fixed the chemical potentials \eqref{Afshar-et-al-gauge} and are thus using the canonical ensemble on the left hand side of \eqref{duality-statement}. [Recall \eqref{bcJ-algebra} and that $\bcJ$ are appropriate near horizon charge operators.] This remarkable feature captured by \eqref{eq:log0} is a consequence of our proposal. Next, we note that $S_{\textrm{\tiny can}}^{{z=0}}=S_{\textrm{\tiny mic}}^{{z=0}}+\frac12 \ln S$ and that $S_{\textrm{\tiny mic}}^{{z=1}}=S^{\textrm{\tiny BTZ}}_{\textrm{\tiny can}}-\frac32 \ln S$ where we used \eqref{eq:log13} and \eqref{eq:log14}.
Putting these together \eqref{eq:log10} implies that (up to terms subleading to log),
\be\label{Fluff=S(z=0)}
S^{z=0}_{\textrm{\tiny mic}}=S^{z=1}_{\textrm{\tiny mic}}-\frac12\ln S =S_{\textrm{\tiny BH}}-2\ln S=S^{\textrm{\tiny fluff}}\;,
\ee
establishing another non-trivial test of our horizon fluff proposal.

\section{Concluding remarks}\label{sec:discussion}

In our concluding section we start with a brief summary of key steps in section \ref{se:6.1}. Then we compare with other approaches that address black hole microstates in section \ref{se:6.2}. Finally, we conclude with selected future research directions in section \ref{se:6.4}.

\subsection{Discussion and summary}\label{se:6.1}
In this work we refined the statement and some details of the horizon fluff proposal put forward in \cite{Afshar:2016uax} and discussed further in \cite{Sheikh-Jabbari:2016npa}. The general picture coming out of our proposal is that black hole microstates, the horizon fluff, belong to a certain subset of near horizon soft hairs that are not distinguishable by the observers away from the horizon. In order to work through our proposal in practice one needs to implement the following three steps: 
\begin{enumerate}
\item Work out the asymptotic symmetry algebra and the corresponding Hilbert space. The black hole is  described by a single state in this Hilbert space. 
\item Work out the near horizon algebra and the associated Hilbert space of soft hairs. Reviewing arguments of \cite{Afshar:2016wfy, Afshar:2016kjj}  we expect to find (copies) of the algebra \eqref{J-algebra} or \eqref{bcJ-algebra} for generic black holes.  
\item Write down a correspondence map [in our case \eqref{duality-statement}] that relates operators/states in the asymptotic Hilbert space and hence the state of the black hole, to operators/states in the Hilbert space of near horizon soft hairs. Solving the equation obtained from this map we get the microstates, the horizon fluff. 
\end{enumerate}

In the present work we implemented all three steps for BTZ black holes.
There are well-established semi-classical frameworks and techniques to work out steps 1.~and 2.  However, in addition to these semi-classical considerations and computations we need to make some `mild' quantum assumptions. For the AdS$_3$ black hole example, these mild quantum assumptions are assuming that the Brown--Henneaux central charge $c$ is quantized and that the conic deficit angle $\nu$ is quantized in units of $1/c$. Both of these assumptions have backing in the cases where we know its full quantum description in the AdS$_3$/CFT$_2$ dualities realized in string theory. Quantization of the deficit angle $\nu$ is also expected by a usual (and rough) intuition: the smallest semi-classical resolution of angles on constant time slice of AdS$_3$ (e.g.~on Poincare disk) is expected to be $\ell_{Pl}/\ell\sim 1/c$ (where $\ell$ is AdS$_3$ radius). 

Regarding the third step and the correspondence map \eqref{duality-statement} we already gave different arguments in its favor. This map was also introduced and used, though with a conceptually different argument by Ba\~nados \cite{Banados:1998wy,Banados:1999ir}. Equation \eqref{duality-statement}  essentially says that the  near horizon and asymptotic energy scales differ by a factor of $1/c$. As an alternative, intuitive argument one may show that in the large $c$ limit this is closely related to the infinite redshift factor at the horizon. This redshift factor becomes finite but large at the stretched horizon \cite{Susskind:1993if} which is at physical distance $\ell_{Pl}$ from the horizon. For the case of AdS$_3$ this ratio becomes $1/c$. More concretely for a generic black hole
\be
\extd s^2=-f(r)\extd t^2+\frac{\extd r^2}{f(r)}+\cdots\,, 
\ee
the generic redshift formula is
\be
{\omega_\infty}=\sqrt{f(r)}\ {\omega_r},
\ee
where $\omega_r$ and $\omega_{\infty}$  are respectively the frequency/energy scale at radius $r$ and the boundary. The near horizon is well described by the Rindler space 
\be\label{Rindler-t-tau}
\extd s^2= -\frac{f'(r_h)^2}{4}\rho^2  \extd t^2+\extd\rho^2+\cdots =-\rho^2 \extd\tau^2+\extd\rho^2+\cdots\,, \qquad \tau= \frac12 f'(r_h) t\,,
\ee
where  black hole temperature $T_{bh}$ is 
\be 
T_{bh}=\frac{f'(r_h)}{4\pi},\qquad r-r_h=\frac{f'(r_h)}{4}\rho^2.
\ee
Now let us put the stretched horizon at physical distance $\ell_{Pl}$ from the horizon, i.e.~it is sitting at coordinate distance $\Delta r_{\textrm{stretched}}$ from the horizon
\be
\Delta r_{\textrm{stretched}}=\ell_{Pl}^2 f'(r_h).
\ee
Therefore, the energy scale at the stretched horizon $\omega_{\textrm{\tiny NH}}$ and the asymptotic one are related as
\bea
\frac{1}{c}\omega_{\textrm{\tiny NH}} (2\pi\ell T_{bh})= \omega_{\infty},
\eea
where  $c$ is the Brown-Henneaux central charge and we took $3\ell_{Pl}=G$. The $2\pi\ell T_{bh}$ factor may be understood as the conversion of units between $\tau$ and $t$ cf.~\eqref{Rindler-t-tau}. The above provides an intuitive reasoning to understand the factor of $1/c$ in \eqref{duality-statement}. 

Each microstate \eqref{BTZ-microstates} can be viewed as the semi-classical limit of a particular full quantum state describing a BTZ black hole with given mass and angular momentum. For future applications it is interesting to further classify microstates into `typical' and `atypical' ones, using standard statistical arguments. For instance, a very atypical microstate is one where all the integers $n_i^\pm$ labelling the microstate
are of order unity $n_i^\pm \sim {\cal O}(1)$. There are very few such microstates. At the other end of the spectrum one can have atypical microstates labelled by an order unity number of integers $n_i^\pm$, so that at least one of them is of order of $c \Delta^\pm$. Again, there are only few such microstates. By contrast, the typical microstates have a large number of integers $n_i^\pm$ most of which are of order of that number. Explicitly, we obtain for typical microstates
\eq{
n_i^\pm\big|_{\textrm{\tiny typical}} \sim {\cal O}(\sqrt{c\Delta^\pm}) \sim
{\cal O}(1)\, c\, |J_0^\pm|\,.
}{eq:typical}
While the motivation for our microstate proposal was geometric, guided by large diffeomorphisms that preserve near horizon boundary conditions, the actual construction of the microstates was purely algebraic. As a consequence, we have no deeper geometric understanding of the `typical' microstates given by \eqref{BTZ-microstates} with \eqref{eq:typical} other than saying that they are a condensate of coherent states of conical defects (see the schematic figure \ref{fig:4}). It is, however, of some interest to consider whether or not some observer can detect these microstates. Even though we have no complete picture yet of dynamical processes like black hole evaporation, based on our algebraic results we can provide the following interpretation. A near horizon observer (someone with access to measuring devices that can, at least in principle, detect correlation functions of the near horizon generators $\cJ_n$) can in principle resolve the precise microstate by performing sufficiently many experiments that effectively act with annihilation operators $\cJ_n$ on the microstate. By contrast, an asymptotic observer (someone with access to either the Virasoro algebra at Brown--Henneaux central charge or the $\hat u(1)$ current algebras generated by $J_n$) has no semi-classical way of resolving the microstates \eqref{BTZ-microstates}. For such an observer a UV completion like string theory seems unavoidable for the purpose of resolving microstates.

The horizon fluff proposal provides a formulation of the intuition and understanding that for a statistical mechanics description of thermodynamical relations we need not have a full knowledge of the underlying quantum theory and some basic knowledge about the presumed microscopic degrees of freedom (microstates) should suffice.  Explicitly, in our horizon fluff proposal, as discussed above, we just used semi-classical results and some basic Bohr-type quantization rather than a detailed UV completed quantum gravity description.\footnote{See also \cite{Hooft:2016cpw} for similar ideas.} Remarkably, within this semi-classical setting we not only derived the Bekenstein--Hawking entropy but also the logarithmic corrections, which are known to provide an IR window to UV information \cite{Sen:2011ba}.

\subsection{Comparison with other approaches}\label{se:6.2}

With the above, we hence arrive at the general picture that the horizon is a like a \textit{fluffball}, covered by the near horizon fluff and that the black hole is in fact a condensate (or solitonic state) of the fluff, which does admit a free (weakly coupled) field description in terms of $\bcW^r$ fields. This free description is a good one only for the soft hair and when we are very close to the horizon. Moving away from the horizon the theory of $\bcW^r$ fluff becomes strongly coupled, where the other picture, the $\bJ$ field description is weakly coupled. This is close to the black hole complementarity picture of Susskind et.~al \cite{Susskind:1993if, Susskind:2005js}. In (an imprecise) sense our proposal is close in spirit to the well-established Strominger--Vafa construction \cite{Strominger:1995cz} where the black hole in string theory (in the weakly coupled regime) is modeled as a bound state of D-branes (recall that D-branes are solitonic states in string theory). There is, however, some important conceptual differences with the Strominger--Vafa setting. Here we do not rely on supersymmetry or (near-)extremality; moreover, our states are labeled by residual diffeomorphism charges that are not distinguishable within a supergravity or string theory setting (which respects the strict notion of the equivalence principle where diffeomorphic geometries are viewed physically equivalent, see \cite{Sheikh-Jabbari:2016lzm} for more discussions).

It is interesting to compare with \cite{Mathur:2011gz}. Starting with the D1-D5 system the usual Brown--Henneaux type of diffeomorphisms were exploited to construct a state space `living at the AdS$_3$-boundary' corresponding to $c=6$, while the remaining states (i.e., nearly all of them) `live' near the region that semi-classically corresponds to the black hole horizon. Those states require some input from complicated UV physics and have a description in terms of fuzzballs (see next paragraph), whereas our approach remains entirely semi-classical in the near horizon description.

Our \emph{fluffball} proposal despite some similarities (including the name) has important conceptual (and also technical) differences with the fuzzball proposal \cite{Mathur:2005zp,Mathur:2008nj,Mathur:2009hf},\footnote{
The fuzzball proposal has been put into test in the D1-D5-P setups. Fuzzball geometries were constructed in supergravity on AdS$_3\times$S$^3\times$T$^4$ or $K_3$. Within this setup, which heavily relies on supersymmetry, the full identification of microstates has been possible only for the so-called two-charge case (see \cite{Chowdhury:2010ct} for more discussions).
} which we now list.
\begin{enumerate}
\item The fuzzball proposal requires an explicit UV completion to construct the microstates, whereas our fluffball is entirely semi-classical. Conceivably, fluffballs could be a `poor man's description' of fuzzballs.
\item As a consequence, there are obvious limitations to our approach that the fuzzball proposal does not have: we will not be able to reliably address very small black holes or the `final flicker' of an evaporating black hole.
\item Conversely, there are several advantages to our approach due to its simplicity; in particular, we could exhaustively list all microstates for non-extremal BTZ black holes, which is a hard problem for fuzzballs and also explain the universality of the Bekenstein--Hawking entropy and its log corrections.
\item In the fuzzball picture the black hole, which is a geometry with horizon, is viewed as a superposition or average of geometries, each of which are smooth and horizon-free \cite{Skenderis:2008qn,Chowdhury:2010ct}. In the fluffball picture the black hole is viewed as a collection of coherent states of particles in AdS$_3$, see the schematic figure \ref{fig:4}.
\item Fuzzballs are not diffeomorphic to each other in any obvious way, while fluffballs differ from each other (and from the vacuum) through soft hair excitations, which corresponds to geometries that are deemed locally identical within a strict GR setting.
\item At this stage the fuzzball proposal offers more on the dynamical aspects of black hole mechanics (including the information paradox). Addressing dynamical aspects is among the open problems to be analyzed in the future within the fluffball proposal.
\end{enumerate}

\subsection{Future research directions}\label{se:6.4}

The statement of our horizon fluff  proposal is based on established frameworks for computing charges for residual gauge symmetries plus some (in our view well-justified) assumptions and proposals that we explained. The first set of future projects should naturally involve providing  further evidence or proof for these assumptions. 
To this end we may need to invoke quantum gravity proposals (e.g.~for quantization of $\nu$ parameter) or further analyze the properties of dual CFT$_2$ and/or conserved charges (for the black hole/particle correspondence we proposed in section \ref{sec:two-desc}). There are also some aspects of our proposal to be studied and explored further. Another line for future works comes from understanding better the nuts and bolts of our current proposal through applying this proposal to other generic black holes.  Below we list some of them that we find more pressing or more interesting:
\begin{itemize}
\item In section \ref{Quant-W-section} we discussed quantization of the primary of weight one operators $\cW$ defined in \eqref{W-nu-def} and \eqref{eq:primary}. From the Chern--Simons formulation of AdS$_3$ gravity  viewpoint we have the $SL(2,\mathbb{R})$ Wilson lines\footnote{Recall that the charges $J(\phi)$ are directly related to $SL(2,\mathbb{R})$ gauge fields of the Chern--Simons theory.} $\boldsymbol{{\cal P}} \text{exp}\left(-2\int^\phi J(\phi)\right)$. On the other hand the commutation relations discussed in section \ref{Quant-W-section} hold for an appropriate ordering.  It is interesting to check if this appropriate ordering is the same as path ordering. 

\item We argued that consistency and closure of the algebra with $\bJ$ operators restricts commutation relation between $\bcW$ fields. We worked to leading order in large $c$. It is desirable to explore this direction further and specify the theory governing $\bcW$ fields. It is also interesting to understand the composite (solitonic) nature of states described by $\bcW$ ($\bJ$) operators from $\bJ$ ($\bcW$) viewpoint. This question is ultimately related to better understanding of the  correspondence proposed in \eqref{duality-statement}. In this regard the observation discussed  in footnote \ref{W-Schawrtz} can be useful.

\item While we motivated the black hole/particle correspondence by showing that there are two different descriptions, both of which lead to a Virasoro algebra at Brown--Henneaux central charge (in one construction this was a classical central charge coming from a twist term, while in the other one it was a large sum of quantum central charges coming from normal ordering of Sugawara terms), we did not prove the correspondence relation \eqref{duality-statement}. Moreover, as stated this correspondence is expected to receive $1/c$ corrections. It is intriguing to explore the possibly of  extending it to a fully fledged duality within the dual CFT$_2$. The correspondence \eqref{duality-statement} imposes strong restriction on the spectrum of the presumed dual CFT$_2$. On the other hand modular invariance also imposes (strong) restrictions on the spectrum of CFT$_2$'s. 
It is hence intriguing to check whether the correspondence implies modular invariance.

\item The black hole/particle correspondence \eqref{duality-statement}, relates the free field limit of the black hole and near horizon soft hair sides. A remarkable fact about this correspondence is that the two sides have a limit where each side admits a free field description. Moreover, as mentioned in section \ref{sec:more-on-HCG}, the closure of the algebra implies that the $\bcW$ sector (the particle side) and the $\bJ$ sector (the black hole sector) should have interactions which are suppressed by a power of $1/c$. Here we restricted ourselves to free field descriptions and deduced kinematical information regarding microstates and their counting.  These interactions are necessary if we want to address more dynamical questions like black hole evaporation or information paradox, which is an obvious direction to pursue in the future.

\item We assumed quantization of the central charge $c=6k$, which semi-classically translates into a quantization of Newton's constant $G$ in terms of the AdS radius, $G^{-1}= 4k/\ell$.  Since we are interested in the large $c$ limit only, this is not a serious loophole in our analysis. While there is no proof that $c$ has to be an integer, it may be argued in the Chern--Simons formulation that the level $k$ (and thus the central charge $c$) should be positive integers. Moreover, modular invariance --- which should be a property of any CFT$_2$ with classical AdS$_3$ gravity dual --- implies that the central charge should be multiple of 24 \cite{Maloney:2007ud} (see also \cite{Gaberdiel:2007ve, Gaiotto:2008jt, Ashrafi:2016mns}).  Note also that conceivably our arguments might generalize to situations in which the central charge is not quantized in positive integers, but some suitable set of positive rational numbers.
\item We postulated quantization of the conical deficit parameter $\nu$ in integers over the central charge, $\nu=r/c$, with $r=1,\dots,c$. This assumption was crucially used in \eqref{J-j-map} to introduce the $\bcJ$ fields, which then in turn facilitated the introduction of the correspondence \eqref{duality-statement}. Quantization of the deficit angle, as we discussed has backing in string setups where it is argued that a resolution of the conic singularity can take place for these values \cite{Maldacena:2000dr,David:2002wn}. 

\item As a consequence of the quantization of $\nu$ and the black hole/particle correspondence,  we learned that the mass of the BTZ black holes should also be quantized: $\Delta^\pm\in \mathbb{Z}/c$.
This feature of our proposal cries for a better understanding and may also be related to restrictions on the spectrum arising from modular invariance of the CFT$_2$.

\item The main ingredient which led us to propose the black hole/particle correspondence was the realization of $\bcW$ fields as weight one primaries with twisted periodicity (cf.~discussions in section \ref{sec:more-on-HCG}). In the footnote \ref{footnote:Fermion-primaries} we pointed out that 
one can construct fermionic primary fields of weight 1/2, $\Psi=e^{{-\int^\phi J}}$ that have periodicity phase  $e^{\mp \pi  \nu i}$. These fermionic fields are anti-periodic for global AdS$_3$ ($\nu=\pm 1$) and periodic for massless BTZ ($\nu=0$). The fermionic fields solve Hill's equation $\Psi''-L\Psi=0$.  The doublet formed out of the two solutions of the Hill's equation form an $SL(2,\mathbb{R})$ doublet \cite{Sheikh-Jabbari:2014nya}, i.e.~fermions in the   Chern--Simons formulation for AdS$_3$ gravity. In this regard the theory of $\Psi$ fields may be relevant to the free limit of the $1+1$ dimensional generalizations of the SYK model \cite{Berkooz:2016cvq,Berkooz:2017efq}. It is interesting to explore if the SYK theory is indeed related to the theory of near horizon soft hairs, and thereby connecting the `black hole type' features uncovered from the SYK theory to the horizon fluff proposal.

\item Our proposal is based on the notion of `near horizon soft hairs' and that when we have a surface of infinite redshift the notion of softness and associated conserved charges can differ at or away from this surface. In our analysis here we focused on black hole horizons; however, our general arguments and analysis should also work for cosmological horizons. We hope the horizon fluff proposal eventually clarifies issues about the de Sitter entropy and thermodynamics.

\item The obvious ultimate target is to construct the Kerr black hole microstates and understanding the microscopic basis of its thermodynamics. While the outline of how the horizon fluff proposal may work for this case was given in \cite{Afshar:2016uax}, it is desirable to work out this proposal in detail. However, before that one may try applying our proposal to other, perhaps simpler, examples of three- or four-dimensional geometries that have event or cosmological horizons.

\end{itemize}
We hope to report on some of the questions and issues discussed above in future publications.

\section*{Acknowledgments}

We thank Glenn Barnich, Marc Henneaux, Finn Larsen, Kyriakos Papadodimas and Ashoke Sen for discussions and Martin Ammon, Steph{\`a}ne Detournay, Wout Merbis, Alfredo Perez, Stefan Prohazka, Max Riegler, David Tempo, Ricardo Troncoso and Raphaela Wutte for collaboration on (near horizon) soft hairy aspects.

HA was supported by Iran's National Elites Foundation (INEF). DG was supported by the Austrian Science Fund (FWF), projects P~27182-N27 and P~28751-N27, and during the final stage also by the program Science without Borders, project CNPq-401180/2014-0. 
The work of HY is supported in part by National Natural Science Foundation of China, Project 11675244.

We also acknowledge the scientific atmosphere of the  workshop on \emph{Quantum Aspects of Black holes}, August 2016, Yerevan Armenia. The work of  MMSh-J is supported in part by junior research chair in black hole physics of the Iranian NSF, the Iranian SarAmadan federation grant and the ICTP network project NT-04. DG and MMSh-J would like to thank the stimulating atmosphere of the focus week \emph{Recent developments in AdS3 black-hole physics} within the program ``New Developments in AdS$_3$/CFT$_2$ Holography'' in April 2017 at GGI in Florence where this work was completed.

\appendix

\section{Ba\~nados geometries and Virasoro Hilbert space}\label{app:A}

AdS$_3$ Einstein gravity is classically defined through the Einstein--Hilbert action
\be\label{AdS3-gravity-theory}
I_{\textrm{\tiny EH}}=\frac{1}{16\pi G}\, \int\extd^3 x\sqrt{|g|}\,\big(R+\frac{2}{\ell^2}\big) + \textrm{boundary\;terms}\qquad
\ee
with the three-dimensional Newton constant $G$ and the AdS$_3$-radius $\ell$. Variation of \eqref{AdS3-gravity-theory} establishes the Einstein equations
\be
 R_{\mu\nu}=-\frac{2}{\ell^2} g_{\mu\nu}\,.
\label{eq:Einstein}
\ee
All solutions to the equations of motion \eqref{eq:Einstein} are locally AdS$_3$, so they are completely specified by the choice of boundary condition. In other words, these solutions are all locally diffeomorphic to each other and can become distinct solutions only through their residual diffeomorphism charges (if they exist). The widely used and studied boundary conditions are the Brown--Henneaux boundary conditions \cite{Brown:1986nw} and the class of solution with Brown--Henneaux boundary conditions are the Ba\~nados geometries \cite{Banados:1998gg} summarized in this appendix, see also \cite{Sheikh-Jabbari:2014nya, Sheikh-Jabbari:2016unm, Compere:2015knw} for more detailed discussions.

\subsection{Locally AdS$_3$ geometries in Ba\~nados coordinate system}
The most general set of locally AdS$_3$  geometries obeying the standard Brown--Henneaux boundary conditions \cite{Brown:1986nw}, are given as \cite{Banados:1998gg}
\be\label{generic-Banados-geometry}
\extd s^2=\ell^2 \frac{\extd r^2}{r^2}- \Big(r\extd y^+- \frac{\ell^2 L_-(y^-)}{r} \extd y^-\Big)\Big(r\extd y^-- \frac{\ell^2 L_+(y^+)}{r} \extd y^+\Big),
\ee
where  $L_\pm(y^\pm)$ are two arbitrary periodic functions $L_\pm (y^\pm+2\pi)=L_\pm(y^\pm)$. We call the metrics \eqref{generic-Banados-geometry} ``Ba\~nados geometries''. The conformal/causal boundary of these geometries is a cylinder parametrised by $t,\phi$, where
$$
y^\pm=t/\ell\pm \phi,\qquad \phi\in[0,2\pi].
$$
As discussed in \cite{Sheikh-Jabbari:2016unm} in principle $r^2$ can take positive or negative values. Some parts in the $r^2<0$ region may be excised due to presence of closed timelike curves.

For general $L_\pm(y^\pm)$, Ba\~nados geometries include black holes with event and Killing horizons and geometries that are not black holes \cite{Sheikh-Jabbari:2016unm}. 
The special case of constant, positive $L_\pm$ constitute the BTZ black hole \cite{Banados:1992wn} family, while constant $-1/4< L_\pm<0$ family are the conic spaces and $L_\pm=-1/4$ is the global AdS$_3$. 

\subsection{Symplectic symmetry algebra}

As discussed in \cite{Compere:2015knw} the family of Ba\~nados geometries form a phase space with a well-defined symplectic two-form on it. The family of transformations under which this symplectic two-form remains invariant is called symplectic symmetries.  
The symplectic symmetry group of the Ba\~nados geometries \eqref{generic-Banados-geometry} is two copies, left and right sector, of Virasoro algebras at the Brown--Henneaux central charge \cite{Compere:2015knw}:
\be\label{B-H-Vir}
\begin{split}
[\bL_n^\pm,\,\bL_m^\pm]&=(n-m)\,\bL^\pm_{n+m}+ \frac{c}{12} \,n^3\, \delta_{n,-m},\\
[\bL_n^+,\bL_m^-]&=0,\qquad  c=\frac{3\ell}{2G}.
\end{split}\ee
The Ba\~nados geometry with $L_\pm(y^\pm)$ is then related to state(s) in the representation of the above algebra such that the expectation value of $\bL_n^\pm$ over those states are specified as
\be\label{Ln-bLn}
L_\pm(y^\pm)=\sum_{n\in\mathbb{Z}} L_n^\pm e^{in y^\pm},\qquad L_n^\pm=\frac{6}{c} \langle \bL_n^\pm\rangle.
\ee
Note that the above expectation value is computed based on the vacuum state where massless BTZ has vanishing $L_0$. The global AdS$_3$ vacuum state which is the $SL(2,\mathbb{R})$ invariant one has $L_0=-1/4$.

\subsection{Virasoro coadjoint orbits, Ba\~nados geometries and Virasoro Hilbert space}\label{Vir-coadj-orbits}

Ba\~nados geometries fall into a one-to-one relation with the coadjoint orbits \cite{Witten:1987ty, Balog:1997zz} of its symplectic symmetries \cite{Sheikh-Jabbari:2016unm}. Since it is relevant to our discussions in this paper  we give a brief account of the Virasoro coadjoint orbits. A more detailed account may be found in \cite{Sheikh-Jabbari:2016unm, Compere:2015knw, Witten:1987ty, Balog:1997zz}. The geometries in Ba\~nados family \eqref{generic-Banados-geometry} specified with $L_\pm$ functions  related through
\be\label{L-Schwarz}
\tilde L(x)=h'^2 L(h(x))-S[h(x);x],\quad h(x+2\pi)=h(x)+2\pi,\quad S[h(x);x]=\frac{h'''}{2h'}-\frac{3h''^2}{4h'^2},
\ee
for any function $h$, are said to be in the same Virasoro coadjoint orbit. When there is no confusion we will drop the $\pm$ subscript to avoid cluttering.
One can recognize two major classes of orbits:  \emph{constant representative orbits}  for which there exists an $h(x)$ where $\tilde L$ becomes a constant, which will be denoted by $L_0$, and \emph{non-constant representative orbits} for which $L$ cannot be mapped to a constant. There are two kinds of non-constant representative orbits, those which are labeled by only an integer and those labeled by an integer and a real-positive number. While a thorough discussion may be found in \cite{Sheikh-Jabbari:2016unm}, here we 
briefly discuss constant representative orbits which is relevant to our discussions in this paper. These orbits are specified by the value of their representative $L_0$ and fall in four categories:
\bn
\item \textbf{Circular orbits} with $L_0=-\frac{n^2}{4},\ n\in\mathbb{N}$. The corresponding geometries (once considering left and right sectors) are generically conic spaces on multiple covers of AdS$_3$. The special case of $n=1$ (in both left and right sectors) corresponds to global AdS$_3$. 
\item \textbf{Elliptic orbits} with $L_0=-\frac{\nu^2}{4},\ \nu\notin\mathbb{Z}$. If both the left and right sectors are in the elliptic orbits, the corresponding geometry is a conic defect or a conformal descendant thereof. If $0<\nu<1$, the representative geometry is a conic space (a particle) on global AdS$_3$, while if $\nu>1$ we have conic singularities on multiple cover of AdS$_3$. 
\item  \textbf{Hyperbolic orbits} with $L_0=b^2,\ b\in\mathbb{R}$. These correspond to the family of BTZ geometries and their conformal descendants. All the geometries in this class have Killing and event horizons; mass and angular momentum of the corresponding BTZ is specified by values of $b_\pm^2$. The special case of $b_\pm=0$ corresponds to massless BTZ.
\item \textbf{Parabolic orbit} with $L_0=0$. If both of the sectors are in parabolic orbits we are dealing with null-self-dual orbifold of AdS$_3$; if one sector is in parabolic and the other in hyperbolic orbits we are dealing with self-dual AdS$_3$ orbifold. The former may be obtained from the near horizon limit of massless BTZ, while the latter from the near horizon limit of generic extremal BTZ \cite{deBoer:2010ac}. The  $L_\pm=0$ case may correspond to massless BTZ or null-self-dual orbifold; the difference between the two is in their global symmetries. 
\en

\paragraph{Hilbert space of Virasoro unitary representations.} Given the classification of the Virasoro  (coadjoint) orbits one may then construct the associated Hilbert space $\cH_{\textrm{\tiny Vir}}$. Representative of any given orbit may be viewed as the vacuum state or lowest weight state and all the other states in the orbit are constructed by the action of all possible combinations of $\bL_{-n},\ n>1$ on this vacuum state. Explicitly, in general the lowest weight state of a given orbit is specified as
\be
\text{Lowest weight (representative) state:}\quad  |(n,L_0)\rangle_X, \qquad n\in \mathbb{Z},\ L_0\in\mathbb{R},
\ee
where $X$ denotes category of the orbit, circular, elliptic, hyperbolic or parabolic. The most general state in the orbit is then
\be\label{HL-generic-state}
 |{{\lbL}}(\{p_i\}); (n,L_0)\rangle_X = {\cal N}_{{\bL}}(\prod_{\{p_i>1\}}\!\!\bL_{-p_i}) |(n,L_0)\rangle_X, \qquad \forall
 |{{\lbL}}(\{p_i\}); (n,L_0)\rangle_X \in {\cal H}_{\textrm{\tiny Vir}},
\ee
where ${\cal N}_{{\bL}}$ is an appropriate normalization factor. 
The full Hilbert space of Ba\~nados geometries is then $\cH_{\text{Ba\~nados}}=\cH_{\textrm{\tiny Vir}}^+\otimes \cH_{\textrm{\tiny Vir}}^-$. In this work, we deal with constant representative orbits, which correspond to $|(0,L_0)\rangle_X$ states and their conformal family/orbit.

The set of Ba\~nados geometries form a phase space with Virasoro algebra \eqref{B-H-Vir} at Brown--Henneaux central charge as its symplectic (and hence, also asymptotic) symmetry algebra. 
By ``Virasoro Hilbert space'' $\cH_{\textrm{\tiny Vir}}$, hereafter and elsewhere in the main text, we mean the set of all Virasoro coadjoint orbits whose states have a well-defined norm on the single cover of AdS$_3$. These are all unitary representation of the two-dimensional conformal group, i.e., primary operators with $\Delta\geq -c/24$  and their conformal descendants. The corresponding geometries are BTZ black holes, conic spaces and global AdS$_3$ and their conformal descendants. Explicitly, the Virasoro Hilbert space decomposes as
\be\label{Hvir}
\cH_{\textrm{\tiny Vir}}=\HBH\cup \HConic\cup \Hglobal\,.
\ee
In terms of Virasoro coadjoint orbits, these are (respectively) hyperbolic, elliptic and circular orbits:
\bi\item {$\HBH$} whose states are labelled by $|\{n_i\};J_0\rangle$ where $|0; J_0\rangle,\ J_0\in \mathbb{R}^+$ denotes the corresponding BTZ state and $\{n_i\}$ its generic conformal (Virasoro) excitations. 
\item
{$\HConic$} whose states are labelled by $|\{n_i\};\nu\rangle$ with $0<\nu<1$.  $|0; \nu\rangle$ denotes the corresponding particle on (global) AdS$_3$  and $\{n_i\}$ its generic conformal (Virasoro) excitations. 
\item {$\Hglobal$} whose states are labelled by $|\{n_i\};NS\rangle$ where $|0; NS\rangle$ denotes the global AdS$_3$ geometry and corresponds to $J_0=i/2$ or $L_0=-1/4$ state. This vacuum state is the $NS$ vacuum of the corresponding dual CFT$_2$ \cite{Maldacena:2000dr, David:2002wn}. In this notation the massless BTZ state $|0;J_0=0\rangle$ is the $R$ vacuum. The $\{n_i\}$ shows generic conformal (Virasoro) excitations. 
\ei
These unitary Virasoro coadjoint orbits contains states for which $L_0$, the expectation value of average of $\bL$, $L_0=\frac{1}{2\pi}\int d\phi \langle \bL(\phi)\rangle$, is bigger or equal to the value of $L_0$ for the representative state \cite{Balog:1997zz}. These orbits are depicted in the left figure of \ref{Hvir-L-J}.

\section{Conic spaces, their metric, symplectic charges and Hilbert space}\label{app:Conic}

Given the importance of conic spaces for our horizon fluff proposal and for completeness, here we briefly discuss some aspects of the conic space geometries. More discussions may be found in \cite{David:2002wn,Sheikh-Jabbari:2016unm,Castro:2011iw}.
\subsection{Particles on AdS$_3$} Let us start with the simple example of particles on AdS$_3$ with given mass and angular momentum. These geometries may be obtained from BTZ black holes in the usual BTZ coordinates via an analytic continuation of the parameters. 
This is the same as taking real $J_0$ to imaginary $J_0$ values (cf.~discussions in sections \ref{sec:3-2} and appendix \ref{app:A}). Explicitly, consider
 \begin{align}
 \extd s^2&=-f(r)\extd t^2+\frac{\extd r^2}{f(r)}+r^2\left(\extd\varphi -\frac{r_+r_-}{\ell r^2}\extd t\right)^2\,,\qquad f(r)=\frac{(r^2-r_+^2)(r^2-r_-^2)}{\ell^2r^2}\nn\\
 &=-\left(\frac{r^2}{\ell^2}-2M\ell\right)\extd t^2+\frac{\extd r^2}{f(r)}+r^2\extd\varphi^2-2J\ell\extd\varphi \extd t\,.
 \end{align}
with $M\ell=\frac{r_+^2+r_-^2}{2\ell^2}$ and $J=\frac{r_+r_-}{\ell^2}$.  One can then recover geometries with conic (or surplus) singularities by the following analytic continuation
\be\label{BH-to-conic}
r_\pm \rightarrow i\ell \rho_\pm\,.
\ee
After changing variable as $\rho^2=(r^2/\ell^2+\rho_-^2)/(\rho_+^2-\rho_-^2)$ the resulting geometry is 
\begin{align}
\extd s^2&=-\left({\rho^2}+\frac{\rho_+^2}{\rho_+^2-\rho_-^2}\right)\extd\tau^2+\ell^2\frac{\extd\rho^2}{\rho^2+1}+\ell^2\left(\rho^2-\frac{\rho_-^2}{\rho_+^2-\rho_-^2}\right)\extd\phi^2+\frac{2\ell\rho_+\rho_-}{\rho_+^2-\rho_-^2}\extd\tau \extd\phi\nn\\
&=\ell^2\frac{\extd\rho^2}{\rho^2+1}-\left({\rho^2}+1\right)\left(\rho_+\,\extd t-\ell\rho_-\extd\varphi\right)^2+\rho^2\left(\rho_+ \ell \extd\varphi-\rho_-\,\extd t\right)^2\label{conic-deficit-AdS-particle}
\end{align}
where $\tau=t\sqrt{\rho_+^2-\rho_-^2}$ and  $\phi=\varphi\sqrt{\rho_+^2-\rho_-^2}$. While $\varphi$ coordinate here has $2\pi$ periodicity, $\phi$ in general has $2\pi\sqrt{\rho_+^2-\rho_-^2}$. We would like to emphasize that there is a one parameter family of global AdS$_3$ spaces, for which $\rho_+^2-\rho_-^2=1$. As one may see from the above metric these cases may be conveniently called global AdS$_3$ in rotating frame. 
The mass and angular momentum of the conic deficit is given by
\bea
\Delta_\pm=(M\ell\pm J)/2, \qquad  \Delta_\pm=-\frac{\nu^2_\pm}{4}\quad\text{where}\quad\nu_\pm=\rho_+\pm\rho_-\,.
\eea
Note that here we are assuming that $0\leq \nu_-\leq \nu_+\leq 1$. This choice is covering 1/4 of the parameter space, the other three sectors are essentially physically equivalent to this sector.

 \subsection{Conic spaces in Chern--Simons formulation}
In section \ref{constant-rep} we assumed that the computation and the algebra of charges for the  family of geometries associated with Virasoro descendants of conic spaces goes through in the same way as the one for black holes. Due to presence of conic deficit in the former and the horizon (with fixed temperature) in the latter, it is desirable to directly work through the computation of the charge for the conic space family. To this end, one uses the Chern--Simons description, following the methods and notation employed in \cite{Afshar:2016wfy, Afshar:2016kjj} --- for related conventions see section \ref{convention}. In an appropriately chosen gauge, the $sl(2,\mathbb{R})_+\times sl(2,\mathbb{R})_-$ gauge fields associated with conic space family is given as in \eqref{eq:bc} of section \ref{convention} (to avoid cluttering we have suppressed $\pm$ index on the gauge fields of  $sl(2,\mathbb{R})_\pm$ gauge groups and concentrate on one sector). 

Let us study the holonomy of the gauge field associated with conic space family which has 
\be\label{nu-J}
{\cal A}=\zeta \extd t+ J \extd\vp, \qquad \int\limits_0^{2\pi} J \extd\varphi=\pi i\nu,
\ee
around the cycle $S^1$ with $\varphi\in[0,2\pi]$ 
\be
\text{Hol}_\varphi(A)={\mathcal P} \exp\left(\oint_{S^1}A\right)=\exp\left(\int\limits_0^{2\pi} A_\varphi \extd\varphi\right)=b^{-1}e^{2\pi i\nu  \bL_0}b=b^{-1}\left(\begin{array}{cc} e^{-\pi i\nu} & 0\\ 0& e^{\pi i\nu}\end{array}\right)b\,.
\ee
The $\text{Hol}_\varphi(A)$ in the Chern--Simons description, as the above equation explicitly shows, determines the $SL(2,\mathbb{R})$ monodromy used in classification of Virasoro coadjoint orbits \cite{Balog:1997zz}.
Considering a circle with $\phi\in[0,2\pi\nu]$ yields a trivial holonomy. 
Therefore, the above holonomy explicitly reflects the nature of the conic deficit and also the periodicity condition on the allowed gauge transformations. We will use the latter to compute the charges. (Here we also do not consider the holonomy along the Euclidean time, as we are not dealing with black holes here.)

The resulting gauge connection associated to conic spaces and its preserving gauge transformation $\delta_\epsilon A= \extd\epsilon +[A,\epsilon]$ are,
\begin{align}
	\delta {\cal A}=\,\delta J(\varphi)\extd\varphi\,,\qquad	\epsilon=\bL_0 \,\eta(\varphi)\,,
\end{align} 
where we have assumed $\delta\zeta=\delta\nu=0$ [where $\nu$ is defined in \eqref{nu-J}] and consequently $\delta_\eta J(\varphi)=\eta'(\varphi)/2$. The $\delta\nu=0$ condition comes from the fact that the charges are ought to be computed in a sector with a given holonomy.
The canonical charges associated with the Chern--Simons theory is \cite{Afshar:2016wfy, Afshar:2016kjj},
\begin{align}
\delta Q[\epsilon]=\frac{k}{2\pi}\oint_{S^1}\llangle\epsilon\,\delta A\rrangle
=\frac{k}{2\pi}\int_0^{2\pi} \extd\varphi\, \delta J(\varphi)\eta(\varphi)\,.
\end{align}
The above yields the algebra \eqref{J-algebra} where now $\bJ_0$ is anti-hermitian and takes imaginary expectation values. 

\subsection{Hilbert space for conic and global AdS$_3$ spaces}

It is straightforward to see that the charge algebra analysis of \cite{Afshar:2016wfy} carries over after the analytic extension \eqref{BH-to-conic}. For the case of conic spaces we get the same algebra as \eqref{J-algebra}, but now $\bJ_0$ is anti-hermitian with eigenvalues $\pm i\nu/2,\ \nu\in(0,1)$. The case of $\nu=1$ corresponds to global AdS$_3$. 

One may construct Virasoro algebra generators as in \eqref{J-L} or \eqref{twisted-sugawara}, however, now $\int_0^{2\pi} L(\phi)$ can take negative values. The associated Hilbert space of unitary representations may be constructed along the same lines discussed in section \ref{sec:3-2}. States in the Hilbert space of conic spaces $\HConic$ can hence conveniently be denoted by $\nu$ and a collection of integers $\{n_i\}$, that is $|\{n_i\}; \nu\rangle$. 

In a similar manner, one can construct the algebra and Hilbert space of global AdS$_3$, which may be denoted as $\Hglobal$. The union of $\Hglobal$ and $\HConic$ forms $\HCG$, cf.~section \ref{sec:3-2}.

\section{Hermitian conjugation and unitarity}\label{appendix-D}

\subsection{$\hat u(1)_k$ current algebra and hermitian conjugation}

The $\hat u(1)_k$ current algebra \eqref{J-algebra} permits different options for hermitian conjugation, namely $\bJ_n^\dagger=\sigma\bJ_{-n}$ for $n\neq0$ with $\sigma^2=1$ and $\bJ_0^\dagger=\sigma_0\bJ_0+\beta$ with $\sigma_0$ and $\beta$ undetermined.  Since the generators $\bL_n$ are twisted Sugawara constructions of $\bJ_n$'s \eqref{twisted-sugawara}, we determine their hermitian conjugation in terms of $\bJ_n^\dagger$:
\begin{align}
\bL_0^\dagger&= \bL_{0}+\tfrac{6}{c}(\sigma_0^2-1)\bJ_0^2 + \tfrac{12}{c}  (\sigma_0\beta\bJ_0+\tfrac{\beta^2}{2} )\\
\bL_n^\dagger&= \bL_{-n} + \tfrac{12}{c} (\sigma\sigma_0-1 )\bJ_0\bJ_{-n}+ \tfrac{12}{c} \sigma\beta\bJ_{-n}-(\sigma -1 )in\bJ_{-n}\;\qquad n\neq0
\end{align}

We always assume hermiticity of the zero-mode Virasoro generator, $\bL_0^\dagger=\bL_0$, which implies $\sigma_0^2=1$ and $\beta=0$. (In the other case, $\beta\neq0$, the latter plays the role of a background charge so that the norm has to be defined as $\langle \sigma_0J_0+\beta|J_0\rangle=1$.) For vanishing $\beta$ we have the following four scenarios for the hermitian conjugate of the symmetry generators on the plane,
\begin{align}
\label{Hermitian-conj}
\sigma&=\sigma_0=1\,,\quad\text{and}\qquad\bL_n^\dagger=\bL_{-n}\quad n\neq0\,,\\
\sigma&=\sigma_0=-1\,,\quad\text{and}\qquad\bL_n^\dagger=\bL_{-n}+2in\bJ_{-n}\quad n\neq0\,,\\
\sigma&=-\sigma_0=1\,,\quad\text{and}\qquad\bL_n^\dagger=\bL_{-n}-\tfrac{24}{c}\bJ_0\bJ_{-n}\quad n\neq0\,,\\
\sigma&=-\sigma_0=-1\,,\quad\text{and}\qquad\bL_n^\dagger=\bL_{-n}-\tfrac{24}{c}\bJ_0\bJ_{-n}+2in\bJ_{-n}\quad n\neq0\,.
\end{align}
In conclusion, assuming hemiticity of the zero mode Virasoro generator $\bL_0$ (implying $\sigma_0^2=1$ and $\beta=0$) and (anti-)hermiticity of $\bJ_n$'s (implying $\sigma^2=1$), the hermitian conjugates of $\bL_{n\neq0}$ are fixed as above. Note that only in the first case the hermitian conjugate of the Virsoro generators $\bL_n$ can be expressed solely in terms of Virasoro generators, while the other three cases also require combinations of current algebra generators.

\subsection{Unitary highest weight representations}
For unitary theories, the norm of all states has to be non-negative. Let us consider a Virasoro primary state $|\phi\rangle$ which is also highest weight state with respect to the current algebra \eqref{J-algebra}. We thus have,
\eq{
\bL_0|\phi\rangle=h|\phi\rangle\,,\quad \bL_{n>0}|\phi\rangle=0\qquad\text{and}\qquad \bJ_0|\phi\rangle=\tfrac{c}{6}J_0|\phi\rangle\,,\quad \bJ_{n>0}|\phi\rangle=0
}{eq:withoutlabel}
where $h=\frac{c}{6}J_0^2$. Using the hermitian conjugates obtained as in \eqref{Hermitian-conj}, the norm of the state $\bL_{-1}|\phi\rangle$ can be computed using the algebras \eqref{J-algebra}, \eqref{L-J-algebra} which are defined on the cylinder as follows.
\begin{align}
\sigma&=\sigma_0=1\,,& ||\bL_{-1}|\phi\rangle||^2&=\langle\phi|\bL_1\bL_{-1}|\phi\rangle=\tfrac{c}{3}\left(J_0^2+\tfrac{1}{4}\right)\langle\phi|\phi\rangle\\
\sigma&=\sigma_0=-1\,,& ||\bL_{-1}|\phi\rangle||^2&=\langle\phi|(\bL_1+2i\bJ_1)\bL_{-1}|\phi\rangle=-\tfrac{c}{12}\left(\nu+1\right)^2\langle\phi|\phi\rangle\\
\sigma&=-\sigma_0=1\,,& ||\bL_{-1}|\phi\rangle||^2&=\langle\phi|(\bL_1-\tfrac{24}{c}\bJ_0\bJ_1)\bL_{-1}|\phi\rangle=\tfrac{c}{12}\left(\nu+1\right)^2\langle\phi|\phi\rangle\\
\sigma&=-\sigma_0=-1\,,& ||\bL_{-1}|\phi\rangle||^2&=\langle\phi|(\bL_1+2i\bJ_1-\tfrac{24}{c}\bJ_0\bJ_1)\bL_{-1}|\phi\rangle=-\tfrac{c}{3}\left(\left(J_0-i\right)^2+\tfrac{1}{4}\right)\langle\phi|\phi\rangle
\end{align}
In cases where $\sigma_0=-1$ we used $J_0=i\nu/2$. The last option $\sigma=-\sigma_0=-1$ obviously has no chance of being unitary. In the second and the third case this state is null if $\nu=-1$. We are specifically interested in the case $\sigma=-\sigma_0=1$ in which the norm is positive definite for $\nu\neq-1$. This algorithm can be continued for the rest of higher level states in the Verma module in terms of the Kac determinants. 

On the plane we have
\eq{
(\bL_n)_{\textrm{\tiny cyl}}=(\bL_n)_{\textrm{\tiny plane}}-\frac{c}{24}\delta_{n,0}\qquad\text{and}\qquad(\bJ_n)_{\textrm{\tiny cyl}}=(\bJ_n)_{\textrm{\tiny plane}}+\frac{ic}{12}\delta_{n,0}
}{eq:sowhat}
and the conformal algebra on the plane becomes
\be\label{L-J-algebra-plane}
[\bL_n,\,\bL_m]=(n-m)\bL_{n+m}+\frac{c}{12}\,(n^3-n)\,\delta_{n,-m}\qquad
[\bL_n,\,\bJ_m]=-m\,\bJ_{n+m}+ \frac{ic}{12} \,(n^2+n)\, \delta_{n,-m}\,.
\ee
This is consistent with the fact that the twisted Sugawara construction on the plane is
\eq{
\bL_n\equiv \frac6{c}\sum_{p\in\mathbb{Z}} \!\bJ_{n-p}\bJ_p + i(n+1)\bJ_n\,.
}{twisted-sugawara2}


\section{$\bcJ,\bcL$-algebras and proof of the Ba\~nados map}\label{app:Banados-map}

In section \ref{sec:more-on-HCG} we introduced the $\bcJ_n$ generators obeying the $\hat u(1)_k$ algebra \eqref{bcJ-algebra}. The generators may be viewed as creation-annihilation operators for a free two-dimensional boson theory on $\mathbb{R}\times S^1$ which is a CFT$_2$. One may then construct the corresponding Virasoro generators  $\bcL_n$'s
\be\label{NH-Vir-gen}
\bcL_n^\pm\equiv\sum_{p\in\mathbb{Z}}\colon\!\bcJ_{n-p}^\pm\,\bcJ_p^\pm\colon\,,
\ee 
where $\colon\colon$ denotes normal ordering ($\bcJ_{-n}=\bcJ_n^\dagger$ and $\bcJ_n, n>0$ are the annihilation operators) and using \eqref{bcJ-algebra} one may readily show that
\be\label{NH-algebra2}
[\bcL_n^\pm,\bcL_m^\pm]=(n-m)\bcL_{n+m}^\pm+\frac{1}{12}n(n^2-1)\delta_{n,-m}\,,\qquad
 [\bcL_n^\pm,\bcJ_m^\pm]=-m \bcJ_{n+m}^\pm\,,
\ee 
and $[\boldsymbol{{\cal X}}^+,\boldsymbol{{\cal Y}}^-]=0$ for any $\boldsymbol{{\cal X}}, \boldsymbol{{\cal Y}}$.  The near horizon algebra consists hence of (two copies of) the Virasoro algebra at central charge one, plus a $\hat u(1)$ current. The vacuum state $|0\rangle$ defined through $\bcJ_n|0\rangle=0,\ n\geq 0$, then satisfies $\bcL_n|0\rangle=0, n\geq-1$ and is hence the $SL(2,\mathbb{R})$ invariant state of the free boson CFT$_2$.

In the Sugawara construction \eqref{Ln-sum-Ln-r} we introduced generators of Virasoro algebra at charge $c$ which recalling the identification \eqref{J-j-map}, one can rewrite them as
\be
\bL_n=\sum_{r=1}^{c}\bL_n^r= \frac{1}{c} (\bcL_{nc}-\frac{1}{24}\delta_{n,0})\,.
\ee
This is indeed the map discussed in \cite{Banados:1998wy} for relating two Virasoro algebras, one at central charge $c$ and the other at central charge one. The $\bcL_n$ algebra includes an additional set of  generators $\bcL_{nc+r}$. Inclusion of these operators will reduce the central charge from $c$ to one. We comment that unlike the $\bcW^r_n$'s and $\bcJ_n$, which are related through the identification \eqref{J-j-map}, $\bL_n^r$ are not related to $\bcL_{nc+r}$; the former for any given $r$ are Fourier modes of periodic functions while the latter are Fourier modes of fields with twisted perioidicty. In a sense our analysis here provides a different viewpoint on the discussions of Ba\~nados in \cite{Banados:1998wy}. Moreover, our discussion clarifies how the Hilbert space of states for the CFT$_2$ discussed above, i.e., the set of parabolic orbits of the Virasoro algebra of $\bcL_n$ is in one-to-one correspondence with $\cH_{\textrm{\tiny CG}}$, the Hilbert space of the Virasoro algebra at central charge $c$.

\providecommand{\href}[2]{#2}\begingroup\raggedright\endgroup

\end{document}